# Scale-dependent Rayleigh-Taylor dynamics with variable acceleration by group theory approach

Snezhana I. Abarzhi*, Kurt C. Williams

Department of Mathematics and Statistics
The University of Western Australia
35 Stirling Highway, Crawley, Perth, WA 60069, Australia

*corresponding author, snezhana.abarzhi@gmail.com

Rayleigh-Taylor instability (RTI) has critical importance for a broad range of plasma processes, from supernovae to fusion. In most instances RTI is driven by variable acceleration whereas the bulk of existing studies have considered only constant and impulsive acceleration. This work focuses on RTI driven by acceleration with power-law time-dependence. We review the existing theoretical approaches, apply the group theory approach to solve this long-standing problem, and yield the unified framework for the scale-dependent dynamics of Rayleigh-Taylor (RT) bubbles and RT spikes. For the early-time linear dynamics we provide the dependence of RTI evolution on the acceleration parameters and the initial conditions. For the late-time nonlinear dynamics, we find a continuous family of asymptotic solutions, directly link the interface velocity to the interface morphology and the interfacial shear, derive solutions for the regular bubbles and for the singular spikes, and study stability of these solutions. The properties of special nonlinear solutions in the RT family are scrupulously described, including the critical, the Taylor, the Layzer-drag, and the Atwood solutions. It is shown that the fastest Atwood bubble is regular and stable, and the fastest Atwood spike is singular and unstable. The essentially multi-scale and interfacial character of RT dynamics is demonstrated. The former can be understood by viewing RT coherent structure of bubbles and spikes as a standing wave with the growing amplitude. The latter implies that RT flow has effectively no motion of the fluids away from the interface and has intense motion of the fluids near the interface, with shear-driven vortical structures appearing at the interface. Our theory agrees with available observations, and elaborates extensive benchmarks for future research and for better understanding of RT driven phenomena in plasmas.

## Section 1 – Introduction

Rayleigh-Taylor instability (RTI) has critical importance for a broad range of processes in nature and technology at astrophysical and at molecular scales [1-4]. Examples of plasma processes governed by RTI include: the fingering of interstellar media along the edges of black holes, the abundance of chemical elements in supernova remnants, the effect of plasma irregularities in the Earth ionosphere on regional climate change, the formation of hot spot in inertial confinement fusion, and the efficiency of plasma thrusters [1-9]. While realistic plasmas are necessarily electro-dynamic, with magnetic fields and charged particles involved, the grip and control of Rayleigh-Taylor (RT) plasma phenomena often requires the better understanding of hydrodynamic aspects of non-equilibrium dynamics and the development of rigorous yet practical theory of RTI in neutral fluids at continuous scales [1,2]. In this work we study the long-standing problem of Rayleigh-Taylor instability driven by acceleration varying as power-law with time, and we employ group theory approach to solve the boundary value problem for the scale-dependent RT dynamics in the early-time linear and the late-time nonlinear regimes [1-4,10-12].

The paper is organized as follows. We start with the brief Introduction in Section 1. We provide the Method in Section 2, and we overview the governing equations (2.1), the theoretical approaches (2.2), and the group theory approach (2.3), including the space groups, the expansions and the dynamical system. We present the Results for the scale-dependent Rayleigh-Taylor instability driven by variable acceleration in Section 3. This includes the early-time linear dynamics (3.1), the late-time nonlinear dynamics (3.2), the focused analysis of the solutions for nonlinear RT bubbles (3.3) and for nonlinear RT spikes (3.4), as well as the properties of RT dynamics with variable acceleration (3.5). We finalize the work with Discussion and conclusion in Section 4, and provide Acknowledgements, Data availability, Author's contributions, References, Tables, and Figure captions and Figures in Sections 5-10.

Rayleigh-Taylor instability develops when fluids of different densities are accelerated against their density gradients, and/or when the fluid interface moves with an acceleration directed toward the denser fluid and normal to the nearly planar interface [3,4]. In the vastly different physical circumstances, at astrophysical and at molecular scales, RT flows have a number of similar features of their evolution [11,12]. RTI starts to develop when the flow fields and the interface are slightly perturbed near the equilibrium state [3,4]. Small perturbations grow quickly (e.g., exponentially in time), and the flow transits to a nonlinear stage, where the growth-rate slows (to a power-law in time). The interface is transformed to a composition of a large-scale structure of bubbles and spikes and small-scale vortical structures [11,12]. The bubbles (spikes) are portions of the light (heavy) fluid penetrating the heavy (light) fluid. Their large-scale dynamics is coherent - as is set by the initial conditions. The small-scale vortical structures are produced by shear at the interface; their dynamics can be irregular. Due to the enhanced interactions and the coupling of scales, RT dynamics becomes complicated. Intense interfacial fluid mixing ensues with time [11-21]. The dynamics of Rayleigh-Taylor mixing is believed to be self-similar. Particularly, for a constant driving acceleration the flow length scale in the acceleration direction grows quadratic in time [11,14-21].

RTI and RT mixing are extremely challenging to study in theory, experiments, and simulations [1,2]. RT experiments use advanced technologies to meet tight requirements for the flow implementation, diagnostics and control [13-18]. RT simulations employ highly accurate numerical methods and massive computations to track unstable interfaces, capture small-scale processes and enable large span of scales [19-21]. In theory we have to reliably grasp non-equilibrium RT dynamics, identify universal properties of the asymptotic solutions, and capture the symmetries of the unstable flow [11,12]. Significant success was achieved recently in the understanding of RTI and RT mixing with constant acceleration [1,2]. In particular, group theory approach [11,12] found the multi-scale character of nonlinear RTI and the order in RT mixing, and explained a broad set of experiments in plasmas and fluids which observed that RT mixing may depart from scenario of canonical turbulence and may keep order at high Reynolds numbers up to 3.2 million [13-18].

In realistic circumstances RT relevant phenomena are often driven by variable acceleration [1,2,5-18,22]: In supernova blast the acceleration is induced by strong variable shocks and RT mixing with variable acceleration enables the conditions for the synthesis of heavy and intermediate mass elements. In inertial confinement fusion, the shocks' induced accelerations and decelerations of the ablation front influence the formation of hot spot. In nano-fabrication, the transformation of materials under high strain rate is governed by RTI with variable acceleration.

In this work we focus on RTI subject to variable acceleration, and we consider accelerations with power-law time-dependence [1, 22,23]. On the side of fundamentals, power-law functions are important to study since they may result in new invariant and scaling properties of the dynamics [10,22]. As regards to applications, power-law functions can be used to adjust the acceleration's time-dependence in realistic environments and to ensure practicality of theoretical results [5-9,22]. To solve the long-standing problem of the scale-dependent early-time and late-time RT dynamics with variable acceleration, we apply group theory approach [11,12,22]. This approach worked remarkably well for constant and impulsive accelerations [11-14]. In the case of variable acceleration, it revealed the new mechanism for energy accumulation and transport at small scales in supernovae [22]. Furthermore, group theory approach found the dependence of the dynamics on the exponent of the acceleration's power-law [22]. Particularly, the scale-dependent interfacial dynamics is of Rayleigh-Taylor type and is driven by the acceleration for the acceleration exponents greater than $(-2)$. It is of Richtmyer-Meshkov type and is driven by the initial growth rate for the acceleration exponents smaller than $(-2)$ [22,23].

Here, we consider RTI for a three-dimensional spatially extended periodic flow with hexagonal symmetry in the plane normal to the acceleration, and we solve the boundary value problem involving boundary conditions at the interface and at the outside boundaries of the domain [10-12,22]. For the early-time linear dynamics we provide the dependence of RTI growth-rate on the acceleration's parameters. For the late-time non-linear dynamics, we find a continuous family of asymptotic solutions, directly link the interface dynamics to the interface morphology and the interfacial shear, and derive the solutions for the regular bubbles and for the singular spikes. The family contains the special solutions for the bubbles and for the spikes, including the critical, the convergence limit, the Taylor, the Layer-drag, the Atwood and the flat solutions. The fastest Atwood bubble is regular and stable, whereas the fastest Atwood spike is singular and unstable. We also reveal the interfacial and multi-scale character of the scale-dependent RT dynamics, and analyze the mechanism of transition from the scale-dependent dynamics to the self-similar mixing for RT bubbles and for RT spikes. Our theory agrees qualitatively with available observations, and elaborates extensive quantitative benchmarks for future research [1,2,13-21]. The properties of the scale-dependent RT dynamics found by our theory - the fields of the velocity and pressure, the interplay of interface morphology and shear with the flow acceleration, the interface growth and growth-rate – serve for better understanding of RT driven plasma phenomena in nature and technology [1,2,5-10,17,18].

## Section 2 - Method

### Sub-section 2.1 – Governing equations

RT dynamics of ideal fluids is governed by the conservation of mass, momentum and energy

$$\frac{\partial \rho}{\partial t}+\frac{\partial \rho v_i}{\partial x_i}=0,\quad \frac{\partial \rho v_i}{\partial t}+\sum_{j=1}^{3}\frac{\partial \rho v_i v_j}{\partial x_j}+\frac{\partial P}{\partial x_i}=0,\quad \frac{\partial E}{\partial t}+\frac{\partial (E+P)v_i}{\partial x_i}=0 \qquad (1)$$

with spatial coordinates $(x_1,x_2,x_3)=(x,y,z)$; time $t$; fields of density, velocity, pressure and energy $(\rho,\mathbf{v},P,E)$, and with $E=\rho\left(e+\mathbf{v}^2/2\right)$ and $e$ being specific internal energy [10,22].

We introduce a continuous local scalar function $\theta(x,y,z,t)$, whose derivatives $\dot{\theta}$ and $\nabla\theta$ exist (the dot marks a partial time-derivative). The function $\theta$ is $\theta=0$ at the interface. The heavy (light) fluid is located in the region with $\theta>0$ ($\theta<0$), and its fields are $(\rho,\mathbf{v},P,E)\to(\rho,\mathbf{v},P,E)_{h(l)}$ and are marked with the subscript $h(l)$ [10-12,22,23]. The ratio of the fluids' densities is parameterized by the Atwood number $A=(\rho_h-\rho_l)/(\rho_h+\rho_l)$ [3,4,10-23].

By using the Heaviside step-function $H(\theta)$, we represent the flow fields in the entire domain as $(\rho,\mathbf{v},P,E)=(\rho,\mathbf{v},P,E)_h H(\theta)+(\rho,\mathbf{v},P,E)_l H(-\theta)$ [11,12,22]. At the interface, the balance of fluxes of mass and normal and tangential components of momentum and energy obey the boundary conditions:

$$\left[\tilde{\mathbf{j}}\cdot\mathbf{n}\right]=0,\quad \left[\left(P+\frac{\left(\tilde{\mathbf{j}}\cdot\mathbf{n}\right)^2}{\rho}\right)\mathbf{n}\right]=0,\quad \left[\left(\tilde{\mathbf{j}}\cdot\mathbf{n}\right)\left(\frac{\left(\tilde{\mathbf{j}}\cdot\boldsymbol{\tau}\right)}{\rho}\right)\boldsymbol{\tau}\right]=0,\quad \left[\left(\tilde{\mathbf{j}}\cdot\mathbf{n}\right)\left(W+\frac{\tilde{\mathbf{j}}^2}{2\rho^2}\right)\right]=0 \qquad (2)$$

where the jump of a quantity across the interface is denoted by $[\ldots]$; the normal and tangential unit vectors of the interface are $\mathbf{n}$ and $\boldsymbol{\tau}$ with $\mathbf{n}=\nabla\theta/|\nabla\theta|$ and $(\mathbf{n}\cdot\boldsymbol{\tau})=0$; the mass flux across the moving interface is $\tilde{\mathbf{j}}=\rho\left(\mathbf{n}\dot{\theta}/|\nabla\theta|+\mathbf{v}\right)$; the specific enthalpy is $W=e+P/\rho$.

The boundary conditions Eqs.(2) are derived from the governing equations Eqs.(1) assuming that the mass flux is conserved at the interface, $\left[\tilde{\mathbf{j}}\cdot\mathbf{n}\right]=\left[\tilde{j}_n\right]=0$, with $\tilde{j}_n=\tilde{\mathbf{j}}\cdot\mathbf{n}$. There is the important particular case, when the conserved mass flux is zero at the interface, $\tilde{j}_n\big|_{\theta=0^\pm}=0$. This case corresponds to a so-called contact discontinuity, or a front, and can describe the interface between two immiscible fluids [10-12,22-24]. The condition $\tilde{j}_n\big|_{\theta=0^\pm}=0$ leads to the continuity of normal component of velocity at the interface $[\mathbf{v}\cdot\mathbf{n}]=0$, and transforms the condition for the conservation of normal component of momentum at the interface to the continuity of pressure at the interface, $[P]=0$. For the zero mass flux at the interface, $\tilde{j}_n\big|_{\theta=0^\pm}=0$, the condition of continuity of tangential component of momentum at the interface holds true for an arbitrary jump of the tangential velocity at the interface $[\mathbf{v}\cdot\boldsymbol{\tau}]=arbitrary$, and the condition of continuity of the energy flux at the interface holds true for any jump of the enthalpy at the interface, $[W]=arbitrary$. In full consistency with the classic results [10], in case of zero mass flux at the interface, $\tilde{j}_n\big|_{\theta=0^\pm}=0$, the boundary conditions at the interface are

$$[\mathbf{v}\cdot\mathbf{n}]=0,\quad [P]=0,\quad [\mathbf{v}\cdot\boldsymbol{\tau}]=arbitrary,\quad [W]=arbitrary \quad (3)$$

Per Eqs.(3), the normal component of velocity and the pressure are continuous at the interface, and the tangential velocity component and the enthalpy can be discontinuous at the interface [10-12,22-24].

The outside boundaries have no influence on the dynamics, and the flow is free from mass sources. This leads to:

$$\mathbf{v}\big|_{z\to+\infty}=0,\quad \mathbf{v}\big|_{z\to-\infty}=0 \quad (4)$$

The initial conditions are the initial perturbations of the interface and the flow's fields.

RTI is driven by the acceleration (the body force or the gravity) $\mathbf{g}$, directed from the heavy to the light fluid, $\mathbf{g} = (0,0,-g)$, $g = |\mathbf{g}|$. In the presence of the acceleration $\mathbf{g}$, the pressure is modified as $P \to P - \rho g z$ [10-12,22-24]. The consider accelerations, which are power-law functions of time, $g = G t^{a}$, $t > 0$. Here $a$ is the acceleration exponent, $a \in (-\infty, +\infty)$, and $G$ is the pre-factor, $G > 0$. Their dimensions are $[G] = m/s^{2+a}$ and $[a] = 1$ [10-12,22,23].

We consider RT flow periodic in the plane $(x, y)$ normal to the $z$ direction of the acceleration $\mathbf{g}$; this periodicity is set by the initial conditions [10-12,22,23]. For ideal incompressible fluids, the initial conditions define the length-scale, the time-scale, and the symmetry of the dynamics. For the length-scale, we choose the spatial period (wavelength) $\lambda$ of the initial perturbation, $[\lambda] = m$, or, for convenience, the wavevector $k$, with $[k] = m^{-1}$ and $1/k \sim \lambda$. For given value of the wavevector $k$ and the acceleration parameters $a, G$, there are two natural time-scales: the time-scale $\tau_G = (kG)^{-1/(a+2)}$ set by the acceleration, and the time-scale $\tau_0 = (k v_0)^{-1}$ set by the initial growth-rate $v_0$ at the initial time $t_0$ [22,23]. For the exponent values $a > -2$, these time-scales relate as $\tau_G << \tau_0$, and $\tau_G$ is the smallest time-scale corresponding to the fastest process. Hence, for Rayleigh-Taylor dynamics driven by the acceleration the relevant time-scale is $\tau_G$ [22,23]. For $a < -2$, the time-scales relate as $\tau_0 << \tau_G$, and for Richtmyer-Meshkov dynamics driven by the initial growth-rate, the relevant time-scale is $\tau_0$ [22,23]. Since we consider Rayleigh-Taylor dynamics in this work, we set the time-scale to be $\tau = \tau_G$, $[\tau] = s$, and we study RT evolution for $t > t_0$ [22,23]

**Sub-section 2.2 – Theoretical approaches**

The mathematical problem of Rayleigh-Taylor instability requires one to solve the system of nonlinear partial differential equations in four-dimensional space-time, solve the boundary value problem at the nonlinear freely evolving discontinuity and the outside boundaries, and also solve the ill-posed initial value problem, with account for singularities developing in a finite time Eqs.(1-4) [11,12,22]. This challenging problem can be handled in well-defined theoretical approximations; empirical models may be applied to interpolate experimental and numerical simulations data [11,12,25].

Having started from the classical works of Rayleigh [3] and Taylor [26], significant progress was achieved in the understanding of the early-time RT dynamics in plasmas and in fluids in a broad range of conditions [27-30]. Particularly, in the linear regime, the RTI growth-rate was derived in a continuous approximation as well as through the solution of the boundary value problem Eqs.(1-4) with account of the effects of surface tension, viscosity, mass diffusion, mass ablation, magnetic fields, compressibility, thermal conductivity, stratification, finite size of the domain, and other factors [3,4,11,12,27-30]. For RT flows with various symmetries and with variable acceleration group theory approach was employed [11-14,18,22,23]. Weakly-nonlinear stages of RTI were studied theoretically and empirically with the focus on the effect of initial spectra on RTI growth-rate [30-32].

Nonlinear dynamics of RTI is a long-standing problem, investigated by Davies & Taylor [4], Layzer [33], Birkhoff [34], Garabedian [35]. Group theory approach analyzed a broad class of Rayleigh-Taylor problems, including dynamics of two- and three-dimensional RT flows with various symmetries [11,12,37]. In case of constant acceleration, it reconciled with one another the approaches [4,33-37], and identified the properties of nonlinear RT dynamics, including, e.g., the tendency to keep isotropy in the plane normal to the acceleration, the discontinuity of the dimensional crossover, and the multi-scale character of the dynamics of nonlinear RT bubbles [11,12].

The interactions of multiple scales further trigger transition of the scale-dependent RT dynamics to the self-similar mixing [13-18]. The transition may occur via the growth of horizontal scales of RT coherent structure, as found by the empirical models, including the interpolation, the buoyancy-drag, and the bubble merge models [39-41,47]. It may also occur due to the dominance of the vertical scale, as first found by group theory approach through the momentum model [11,42].

By further assuming that the self-similar RT mixing is similar to canonical turbulence due to its large Reynolds number, the interpolation and turbulence models were developed; they substituted the growth-rate of RT mixing in turbulent scaling laws and computed turbulence effects [25,43-47]. By being harmonious with the theory of canonical turbulence [10,48] and with models [39-41,43-47], and by analyzing symmetries and invariants of RT dynamics, group theory approach identified, through the momentum model, the properties of RT mixing not considered in other studies [11,13,25,42]. Particularly, it found that in case of constant acceleration the self-similar RT mixing may exhibit order, since it has stronger correlations, steeper spectra, weaker fluctuations, and stronger sensitivity to deterministic conditions when compared to canonical turbulence [11,12]. These results explained the high Reynolds number experiments in plasmas and fluids [13-15,17,18].

**Sub-section 2.3 – Group theory approach for Rayleigh-Taylor dynamics**

In this work, we apply group theory approach to yield the unified framework for the scale-dependent RT dynamics with variable acceleration in the early-time linear and the late-time nonlinear regimes [11-14,22,23]. We solve the governing equations Eqs.(1-4) by employing theory of space groups, by using canonical forms of the Fourier series and spatial expansions, and by accounting for non-locality and singularities of the interface dynamics [12,37,38,49]. For the early-time regime, we provide the dependence of RT evolution on the acceleration parameters and the initial conditions. For the late-time regime, we find nonlinear solutions for regular RT bubbles and for singular RT spikes, and provide physics interpretations of mathematical attributes of the boundary value problem Eqs.(1-4). By analyzing properties of regular RT bubbles and singular RT spikes, we further outline the mechanism of the flow transition to the self-similar mixing. We address to the future the establishment of a direct link of the scale-dependent RT dynamics to the self-similar RT mixing within group theory approach.

**Sub-section 2.3.1 – Space groups**

For spatially periodic flows, space groups can be applied, since RT dynamics is invariant with respect to a group $\mathbf{G}$, whose generators are translations in the plane, rotations and reflections [11,12,49]. These groups are also known as Fedorov and/or Schoenflies groups, and they are commonly used in physics [10,49]. While there are 7 one-dimensional and 17 two-dimensional space groups, only some of these groups should be considered. Particularly, groups relevant to structurally stable RT dynamics must have the anisotropy in the acceleration direction and the inversion in the normal plane, such as groups of hexagon $p6mm$ or square $p4mm$ for a three-dimensional (3D) flow, and group $pm11$ in a two-dimensional (2D) flow [10-12,49]. We use international classification of the space groups, where $p$ stands for periodicity in one (two) direction(s), and, for each of spatial directions, 1 stands for unit element, $m$ stands for mirror plane of reflection, and $6(4)$ stands for $6(4)$-fold axis of rotation [10,49].

By using the techniques of group theory, we apply irreducible representations of a relevant group to expand the flow fields as Fourier series. We further make spatial expansions in a vicinity of a regular point at the interface (i.e., the tip of the bubble or the tip of the spike). The governing equations are then reduced to a dynamical system in terms of surface variables and moments. The system's solutions are sought, and their stability is studied [11,12,37,38]. More details on group theory approach can be found in works on RTI with constant acceleration [11,12,14,22,37,38].

**Sub-section 2.3.2 – Fourier series and spatial expansions**

We consider incompressible dynamics with negligible stratification and density variations. For incompressible ideal fluids, the spatial period (wavelength) $\lambda$ of the initial perturbation defines the length-scale of the early-time linear and the late-time nonlinear dynamics. We focus on the large-scale coherent dynamics with length scale $\sim \lambda$, presuming that shear-driven interfacial vortical structures are small with length scales $<< \lambda$ [11,12].

For the large-scale coherent structure, the fluid motion is potential in the bulk, and the velocity of the heavy (light) fluid is $\mathbf{v}_{h(l)} = \nabla\Phi_{h(l)}$. For incompressible ideal fluids the equation for the conservation of mass in Eqs.(1) leads to the Laplace equation $\Delta\Phi_{h(l)} = 0$ for $\theta > 0 (< 0)$ [10-12]. At the outside boundaries of the domain the velocity of the heavy (light) fluid vanishes, $\nabla\Phi_{h(l)}\big|_{z\to+\infty(-\infty)} = (0,0,0)$, Eqs.(4). At the regular point of the interface - the tip of the bubble (spike) - with the coordinate $(0,0,z_0(t))$ the velocity of the heavy (light) fluid is $\nabla\Phi_{h(l)}\big|_{(0,0,z_0)} = (0,0,v(t))$, where $v(t) = \partial z_0/\partial t$.

We consider three-dimensional coherent structure with hexagonal symmetry in the plane normal to the acceleration, Figure 1 [10-12,22,23,49]. In order to make a Fourier series expansions of the flow fields, we recall that for group $p6mm$ with hexagonal lattice the spatial periods $\boldsymbol{a}_{1(2)}$ have equal lengths, $|\boldsymbol{a}_{1(2)}| = \lambda$, and are inclined relative to one another at the angle $2\pi/3$, with $\boldsymbol{a}_1 \cdot \boldsymbol{a}_2 = -1/2$. The wave-vectors $\boldsymbol{k}_{1(2)}$ of the reciprocal lattice are defined by the relations $\boldsymbol{k}_{1(2)} \cdot \boldsymbol{a}_{2(1)} = 0$ and $\boldsymbol{k}_{1(2)} \cdot \boldsymbol{a}_{1(2)} = 2\pi$. The wave-vectors $\boldsymbol{k}_{1(2)}$ are inclined relative to one another at the angle $\pi/3$, with $\boldsymbol{k}_1 \cdot \boldsymbol{k}_2 = 1/2$, and have equal length $k = 4\pi/(\lambda\sqrt{3})$. Linear combination of independent spatial periods $\boldsymbol{a}_{1(2)}$ defines the spatial period $\boldsymbol{a}_3 = -(\boldsymbol{a}_1 + \boldsymbol{a}_2)$, and linear combination of independent wave-vectors of the reciprocal lattice $\boldsymbol{k}_{1(2)}$ defines the wavevector $\boldsymbol{k}_3 = \boldsymbol{k}_1 - \boldsymbol{k}_2$ [11,37,49].

For convenience, we perform the derivations in the non-inertial frame of reference moving with the velocity $v(t)$ in the $z$-direction relative to the laboratory frame of reference [11,12,23,37]. Then, the Fourier series expansions of the potentials $\Phi_{h(l)}$ have the form:

$$\Phi_h(\boldsymbol{r},z,t) = \sum_{m=1}^{\infty} \Phi_m(t) \left( z + \frac{exp(-mkz)}{3mk} \sum_{i=1}^{3} cos(m\boldsymbol{k}_i\boldsymbol{r}) \right) + f_h(t) + cross\ terms,$$
$$\Phi_l(\boldsymbol{r},z,t) = \sum_{m=1}^{\infty} \tilde{\Phi}_m(t) \left( -z + \frac{exp(mkz)}{3mk} \sum_{i=1}^{3} cos(m\boldsymbol{k}_i\boldsymbol{r}) \right) + f_l(t) + cross\ terms \qquad (5)$$

Here $\boldsymbol{r} = (x,y)$, $\Phi_m(\tilde{\Phi}_m)$ are the Fourier amplitudes of the heavy (light) fluid, $f_{h(l)}$ are time-dependent functions, and $m$ is natural number [11,12,22,37]. Cross terms appear in high orders. The Fourier series in Eqs.(5) hold true upon group $p6mm$ transformations, including rotations on angles $\pm\pi/3, \pm 2\pi/3, \pm\pi$ around the $z$- axis, and reflections $x \to -x$, $y \to -y$ and $(x,y) \to -(x,y)$ in the mirror planes along the $x$- and $y$- axes. The Fourier series in Eqs.(5) ensure that in the moving non-inertial frame of reference the tip of the bubble (spike) with coordinates $(0,0,0)$ is the stagnation point with $\nabla\Phi_{h(l)}\big|_{(0,0,0)} = (0,0,0)$, and at the outside boundaries $\nabla\Phi_{h(l)}\big|_{z\to+\infty(-\infty)} = (0,0,-v(t))$.

In the laboratory frame of reference near the tip of the bubble (spike) the interface has the form $Z = z_0(t) + z^*(\mathbf{r},t)$, and the interfacial function $\theta$ is $\theta = -z + Z = -z + z_0(t) + z^*(\boldsymbol{r},t)$. In the moving non-inertial frame of reference near the tip of the bubble (spike) the interface is described as $\overline{Z} = z^*(\boldsymbol{r},t)$ and the function $\theta$ is $\theta = -z + \overline{Z} = -z + z^*(\boldsymbol{r},t)$, where the function $z^*(\boldsymbol{r},t)$ is defined locally as

$$z^* = \sum_{N=1}^{\infty} \zeta_N(t)\,\boldsymbol{r}^{2N} + cross\ terms \quad (6)$$

Here $N$ is natural number, $\zeta_N$ are the surface variables, $\zeta_1 = \zeta$ is the principal curvature at the bubble (spike) tip. Cross terms appear in high orders [11,12,22,23,37].

**Sub-section 2.3.3 – Dynamical system**

Upon substituting these expansions in the governing equations and further expanding the equations in the vicinity of the regular point of the interface – the tip of the bubble (spike) - we derive from the governing equations a dynamical system in terms of moments and surface variables [11,12,22,23,37]. For group $p6mm$, at $N = 1$, the interface is $z^* = \zeta(x^2 + y^2)$, and the dynamical system is:

$$\rho_h\left(\dot{\zeta} - 2\zeta M_1 - \frac{M_2}{4}\right) = 0, \quad \rho_l\left(\dot{\zeta} - 2\zeta\tilde{M}_1 + \frac{\tilde{M}_2}{4}\right) = 0,$$

$$\rho_h\left(\frac{\dot{M}_1}{4} + \zeta\dot{M}_0 - \frac{M_1^2}{8} + \zeta g\right) = \rho_l\left(\frac{\dot{\tilde{M}}_1}{4} - \zeta\dot{\tilde{M}}_0 - \frac{\tilde{M}_1^2}{8} + \zeta g\right),$$

$$-M_1 + \tilde{M}_1 = arbitrary, \quad M_0 = -\tilde{M}_0 = -v, \qquad (7)$$

These equations represent the continuity of the normal component of velocity and the normal component of momentum at the interface, the discontinuity of the tangential component of velocity at the interface, and the absence of mass sources. The moments $M_n(\tilde{M}_n)$ are:

$$M_n = \sum_{m=1}^{\infty} \Phi_n k^n m^n + cross\ terms, \quad \tilde{M}_n = \sum_{m=1}^{\infty} \tilde{\Phi}_n k^n m^n + cross\ terms, \quad n = 0,1,2,... \quad (8)$$

Since the tangential component of the velocity can have a jump at the interface, there can be the interfacial shear Eqs.(3). In order to quantify the interfacial shear, we define the shear function as the directional derivative of the jump of the tangential component of velocity at the interface in the direction of the tangential vector at the interface. In the vicinity of the regular point – the tip of the bubble (spike), at $N = 1$, the shear function is $\Gamma = \Gamma_{x(y)}$, with $\Gamma_{x(y)} = \partial \lfloor v_{x(y)} \rfloor / \partial x(y)$, and $\Gamma = -M_1 + \tilde{M}_1$ in Eqs.(7) [22]. According to the boundary conditions in Eqs.(3) and Eqs.(7), the shear value can be arbitrary.

In dynamical system Eqs.(7), the initial conditions at time $t_0$ are the initial curvature $\zeta_0 = \zeta(t_0)$ and velocity $v(t_0)$; the latter defines the initial growth-rate $v_0 = |v(t_0)|$. As time evolves, bubbles move up and are concave down, and spikes move down and are concave up, Figure 1. For Rayleigh-Taylor dynamics with $a > -2$, we set the length-scale in the system Eqs.(7) to be $1/k$, where the wavevector $k$ is $k = 4\pi/\sqrt{3}\lambda$, and the time-scale to be $\tau = \tau_G = (kG)^{-1/(a+2)}$ [22,23]. We find solutions for RT bubbles and RT spikes in the early time linear regime, $(t - t_0) << \tau$ and in the late time nonlinear regime, $(t - t_0) >> \tau$ with $t >> \tau$ and $t >> t_0$.

**Section 3 – Results**

**Sub-section 3.1 – Early-time linear dynamics**

**Sub-section 3.1.1 – Solutions for linearized system**

For the early-time dynamics with $(t - t_0) << \tau$, only the first order harmonics can be retained in the expressions for the momentum, $M_n = k^n \Phi_1$, $\tilde{M}_n = k^n \tilde{\Phi}_1$, $n = 0,1,2$ with $M_0 = -\tilde{M}_0 = -v$ [23]. For Rayleigh-Taylor dynamics with $a > -2$, for small perturbations the dynamical system is linearized, leading to

$$\dot{\zeta} + \frac{k^2}{4} v = 0, \quad \dot{v} + \frac{4A}{k} \zeta G t^a = 0 \quad (9)$$

Its solution is

$$\frac{\zeta}{k} = C_1 \sqrt{\frac{t}{\tau}}\, I_{\frac{1}{2s}}\left( \frac{\sqrt{A}}{s} \left( \frac{t}{\tau} \right)^s \right) + C_2 \sqrt{\frac{t}{\tau}}\, I_{-\frac{1}{2s}}\left( \frac{\sqrt{A}}{s} \left( \frac{t}{\tau} \right)^s \right), \quad v = -\frac{4}{k} \frac{d}{dt} \frac{\zeta}{k}, \quad s = \frac{a+2}{s} \quad (10)$$

where $I_p$ is the modified Bessel function of the $p$th order, Table 1 [22,23]. For constant acceleration, $a = 0$, this solution reproduces the classical result. Figure 2 illustrates the solutions in Eqs.(10) for the early-time dynamics of RT bubbles and RT spikes at various values of the acceleration exponent $a$. In the linear regime, the growth and the growth-rates of the bubbles and the spikes are nearly symmetric.

Note that Richtmyer-Meshkov dynamics with $a<-2$ is driven by the initial growth-rate $v_0$ and is independent of the acceleration; its early-time solution is

$$\frac{\zeta}{k}=-\frac{1}{2A}ln\left(C_2\frac{t}{\tau_0}+C_1\right),\quad v=-\frac{4}{k}\frac{d}{dt}\frac{\zeta}{k},\quad \tau_0=\frac{1}{kv_0}$$

For $a>-2$ the contribution of these initial-growth-rate driven terms is negligible when compared to those induced by the acceleration in Eqs.(10).

**Sub-section 3.1.2 – Effect of initial conditions**

Solutions for linearized system Eqs.(9) describe, as time evolves, the dynamics of bubbles with $\zeta\leq 0, v\geq 0$ and spikes with $\zeta\geq 0, v\leq 0$. To study how the bubbles and spikes are being formed, we consider the very early time dynamics, $t\approx t_0$ [22,23]. The solution for Eqs.(9) linearized for $t\approx t_0$ is

$$\zeta-\zeta(t_0)\approx-\frac{k}{4}\left(\frac{t-t_0}{\tau}\right)sgn\left(\frac{v(t_0)}{v_0}\right),\quad v-v(t_0)\approx-4A\frac{(\zeta_0/k)^{-1}}{(\tau k)}\left(\frac{t_0}{\tau}\right)^a\left(\frac{t-t_0}{\tau}\right)\quad(11)$$

This suggests the following changes of the morphology and the velocity of the interface near the tip.

For $v(t_0)/v_0>0$ and $\zeta_0 k<0$, the interface becomes more curved and its velocity increases. For $v(t_0)/v_0>0$ and $\zeta_0 k>0$, the interface flattens and the velocity decreases. For $v(t_0)/v_0<0$ and $\zeta_0 k>0$, the interface becomes more curved and the velocity magnitude increases. For $\zeta_0 k<0$ and $v(t_0)/v_0<0$, the interface flattens and the velocity magnitude decreases. For the acceleration-driven Rayleigh-Taylor dynamics with $a>-2$, the bubbles are formed for $\zeta_0 k<0$, the spikes are formed for $\zeta_0 k>0$. The positions of RT bubbles and RT spikes are defined by the initial morphology of the interface, in excellent agreement with observations [11-23].

**Sub-section 3.2 – Late-time nonlinear dynamics**

**Sub-section 3.2.1 - Asymptotic behavior of nonlinear solutions**

For the late-time dynamics, we seek asymptotic solutions in the form of power-law functions

$$\zeta\sim k\left(\frac{t}{\tau}\right)^{\beta},\quad v\sim\frac{1}{k\tau}\left(\frac{t}{\tau}\right)^{\alpha},\quad \Phi_m,\tilde{\Phi}_m\sim\frac{1}{k\tau}\left(\frac{t}{\tau}\right)^{\alpha},\quad M_n,\tilde{M}_n\sim\frac{k^n}{k\tau}\left(\frac{t}{\tau}\right)^{\alpha}\quad(12)$$

where the exponents $\alpha, \beta$ are defined from the dominant balance conditions in system Eqs.(7). By substituting these expressions into the system Eqs.(7), we find that for $a > -2$, the dominant balance conditions require the power-law exponents to be

$$\beta = 0, \quad \alpha = \frac{a}{2} \quad (13)$$

This leads [22] to asymptotic nonlinear solutions with

$$\zeta \sim k, \quad v \sim \frac{1}{k\tau}\left(\frac{t}{\tau}\right)^{a/2}, \quad \Phi_m, \tilde{\Phi}_m \sim \frac{1}{k\tau}\left(\frac{t}{\tau}\right)^{a/2}, \quad M_n, \tilde{M}_n \sim \frac{k^n}{k\tau}\left(\frac{t}{\tau}\right)^{a/2} \quad (14)$$

Note that for Richtmyer-Meshkov dynamics with $a < -2$ the dominant balance conditions require the power-law exponents to be $\beta = 0, \alpha = -1$ and lead to asymptotic nonlinear solution with $\zeta \sim k, v \sim (k\tau)^{-1}(t/\tau)^{-1}, M_n, \tilde{M}_n \sim k^n (k\tau)^{-1}(t/\tau)^{-1}$. For $a > -2$, the contribution of these terms is negligible, and Rayleigh-Taylor nonlinear dynamics is independent of the initial growth-rate.

**Sub-section 3.2.2 – Family of nonlinear asymptotic solutions**

To find for $a > -2$ asymptotic nonlinear solutions with $\zeta \sim k$ and $M_n, \tilde{M}_n \sim k^n (t/\tau)^{a/2}$, we retain higher order harmonics in the expressions for moments $M_n, \tilde{M}_n$. Each of the moments is an infinite sum of the weighted Fourier amplitudes [11,12,22,36-38]. To solve the closure problem, group theory is further applied. Particularly, the nonlinear asymptotic solutions are not unique and form a continuous family. Their multiplicity is due to the singular and non-local character of RT dynamics [35-38]. The number of the family parameters is set by the flow symmetry. For group $p6mm$ the dynamics is highly isotropic, $z^* \sim \zeta(x^2 + y^2)$, the interface morphology is captured by the principal curvature $\zeta$, and the family has one parameter [11,12,37,38]. The multiplicity of the nonlinear asymptotic solutions is also due to the presence of shear $\Gamma$ at the interface Eqs.(3), Eqs.(7) [22].

At $N = 1$, for the family of the nonlinear solutions, the velocity $v$, the shear function $\Gamma$ and the Fourier amplitudes $\Phi_m (\tilde{\Phi}_m)$ depend on the curvature $\zeta$ as (see Table 2 for the explicit expressions):

$$v = \frac{1}{k\tau}\left(\frac{t}{\tau}\right)^{a/2} S\left(A, \frac{\zeta}{k}\right), \quad \Gamma = \frac{1}{\tau}\left(\frac{t}{\tau}\right)^{a/2} Q\left(A, \frac{\zeta}{k}\right) \quad (15)$$

$$\Phi_m = \frac{1}{k\tau}\left(\frac{t}{\tau}\right)^{a/2} \phi_m\left(A, \frac{\zeta}{k}\right), \quad \tilde{\Phi}_m = \frac{1}{k\tau}\left(\frac{t}{\tau}\right)^{a/2} \tilde{\phi}_m\left(A, \frac{\zeta}{k}\right).$$

Our dynamical system Eqs.(7) can account for any number of harmonics $m$ in any order $N$. In higher orders, $N > 1$, the solutions can be found similarly to the $N = 1$ case [11,12,37,38]. Briefly, for $N > 1$ in a broad range of curvature values $(\zeta/k)$ the asymptotic solutions exist and converge with the increase in $N$, the lowest-order amplitudes $|\Phi_1|, |\tilde{\Phi}_1|$ are dominant, and the values of amplitudes $|\Phi_m|, |\tilde{\Phi}_m|$ decay with the increase in $m$. Group theory analysis of RT dynamics with constant acceleration, $a = 0$, find that the $N = 1$ nonlinear solutions properly capture the physical behavior [11,12,37,38]. We address to the future the detailed investigation of mathematical properties of the nonlinear asymptotic solutions in RTI with variable acceleration at $N > 1$.

The nonlinear asymptotic solutions in the family Eqs.(15) describe both the bubbles, which move up and are concave down, and the spikes, which move down and are concave up. Figure 3 illustrate the dependence of the velocity and the shear function on the interface morphology (i.e., the curvature) for the nonlinear bubbles and the nonlinear spikes at various Atwood numbers in the family Eqs.(15). The curvature values are scaled with $\zeta_{cr} = -(3/8)k$ for the bubbles and with $\hat{\zeta}_{cr} = (3/8)k$ for the spikes. While the velocity and the shear function of the asymptotic nonlinear solutions in Eqs.(15) are time-dependent, when scaled with the values of the acceleration $g = Gt^a$ and the wavevector $k$, their magnitudes $\left|v/\sqrt{g/k}\right|$ and $\left|\Gamma/\sqrt{gk}\right|$ depend only on the interface morphology $|\zeta/k|$ and the Atwood number $A$. Nonlinear solutions in Figure 3 belong to the same family Eqs.(15), with $(\zeta/k) < 0$ for bubbles and with $(\zeta/k) > 0$ for spikes. Nonlinear bubbles are regular, with finite values $\left|v/\sqrt{g/k}\right|$ and $\left|\Gamma/\sqrt{gk}\right|$, whereas nonlinear spikes may be singular, with the re-scaled values of the velocity $\left|v/\sqrt{g/k}\right|$ and the interfacial shear $\left|\Gamma/\sqrt{gk}\right|$ both approaching infinity at some finite $(\zeta/k)$, Figure 3.

In the following sub-sections we study in details the properties of bubbles and spikes in the scale-dependent nonlinear dynamics of RTI with variable acceleration.

**Sub-section 3.3 – Dynamics of nonlinear bubbles**

**Sub-section 3.3.1 – Nonlinear solutions**

For nonlinear bubbles, we consider solutions with $v(\tau k) > 0, (\zeta/k) < 0$ in the family Eqs.(15), Table 2. In the family of asymptotic solutions with negative curvatures $\zeta/k < 0$, the velocities are positive, $v(\tau k) > 0$, and the shear functions are positive, $\Gamma\tau > 0$. Nonlinear bubbles move down and are concave up, and vortical structures, which may appear in a vicinity of the bubble tip due to the interfacial shear, 'rotate' from the heavy to the light fluid, in agreement with experiment [14-18].

The solutions for RT bubbles with $(\zeta/k)<0, v(\tau k)>0, \Gamma\tau>0$ in Eqs.(15) have the following properties. The domain of the function $v$ is $\zeta \in (\zeta_{cr}, 0), \zeta_{cr} = -(3/8)k$, where the sub-script stands for critical. The range of the velocity function is $v \in (0, v_{max})$, where the zero value $v=0$ is achieved at $\zeta=0$ and at $\zeta=\zeta_{cr}$, and where the maximum value $v=v_{max}$ is achieved at $\zeta=\zeta_{max}$ with $\zeta_{max} \in (\zeta_{cr}, 0)$. The fastest solution in the family $(\zeta_{max}, v_{max})$ obeys the condition $\partial v/\partial\zeta|_{\zeta=\zeta_{max}}=0$ with $\partial^2 v/\partial\zeta^2|_{\zeta=\zeta_{max}}<0$. The explicit dependence of the curvature $\zeta_{max}$ and the velocity $v_{max}$ on the Atwood number is complex. For $\zeta \in (\zeta_{cr}, 0)$ the shear function $\Gamma$ is 1-1 function on the curvature $\zeta$ with $\Gamma \in (\Gamma_{min}, \Gamma_{max})$. The minimum value $\Gamma_{min}=0$ is achieved at $\zeta=0$, and the maximum value $\Gamma_{max}$ is achieved at $\zeta=\zeta_{cr}$. For $\zeta \in (\zeta_{cr}, 0)$ and $\Gamma \in (0, \Gamma_{max})$, the bubble velocity $v$ is a 1-1 function on the shear $\Gamma$. Explicit functions $v(\Gamma)$ and $\Gamma(v)$ are cumbersome.

For nonlinear RT bubbles, Figure 4a, Figure 4b, Figure 4c illustrate the dependence of the re-scaled velocity $v/v_{max}$ and the re-scaled shear function $\Gamma/\Gamma_{max}$ on the re-scaled curvature $\zeta/\zeta_{cr}$, as well as the dependence of the shear $\Gamma/\Gamma_{max}$ on the velocity $v/v_{max}$ at some Atwood numbers $A$. The universality of nonlinear dynamics of RT bubbles and the link between the interface dynamics, interface morphology and the interfacial shear are clearly seen from these plots. Figure 4d further illustrates the convergence of nonlinear RT bubbles with $\zeta=\zeta_{max}$ by showing that the lowest-order amplitudes $|\Phi_1|, |\tilde{\Phi}_1|$ are dominant for a broad range of the Atwood numbers $A$.

According to our results, the late-time dynamics of nonlinear RT bubbles is regular, in agreement with experiments, Figure 3, Figure 4 [14-18,22].

**Sub-section 3.3.2 – Special solutions**

The family of asymptotic solutions for nonlinear RT bubbles with $\zeta/k<0$ has some special solutions. We call these solutions the critical bubble, the convergence limit bubble, the Taylor bubble, the Layzer-drag bubble, the Atwood bubble, and the flat bubble. Properties of these solutions are illustrated by Figure 5.

A. The critical bubble

For the critical bubble with curvature $\zeta = \zeta_{cr}$ the solution is

$$\zeta_{cr} = -\frac{3}{8}k, \quad v_{cr} = 0, \quad \Gamma_{cr} = \frac{1}{\tau}\left(\frac{t}{\tau}\right)^{a/2}\sqrt{\frac{6A}{1+A}}, \quad \Gamma_{cr} = \Gamma_{max} \quad (16)$$

This solution is special, since the bubbles in the family of solutions cannot be more curved than the critical bubble. For the critical bubble the velocity is zero and the shear function achieves its maximum value. The large shear can alone maintain the pressure at the interface, Figure 5.

B. The convergence limit bubble

At $N = 1$, the amplitudes of the heavy fluid relate as $|\Phi_1| > |\Phi_2|$ for $\zeta \in (\zeta_{cn}, 0)$ and as $|\Phi_1| = |\Phi_2|$ at $\zeta = \zeta_{cn}$, where $\zeta_{cn} = -(5/24)k$ and sub-script stands for convergence. The amplitudes of the light fluid relate as $|\tilde{\Phi}_1| > |\tilde{\Phi}_2|$ for $\zeta \in (\zeta_{cr}, 0)$ and as $|\tilde{\Phi}_1| = |\tilde{\Phi}_2|$ at $\zeta = \zeta_{cr}$. This defines the convergence limit bubble solution.

For the convergence limit bubble with curvature $\zeta = \zeta_{cn}$, the solution is

$$\zeta_{cn} = -\frac{5}{24}k, \quad v_{cn} = \frac{1}{k\tau}\left(\frac{t}{\tau}\right)^{a/2}\frac{14}{3}\sqrt{\frac{10A}{3(45+53A)}}, \quad \Gamma_{cn} = \frac{1}{\tau}\left(\frac{t}{\tau}\right)^{a/2} 9\sqrt{\frac{10A}{3(45+53A)}} \quad (17)$$

At $N = 1$, the convergence holds for $\zeta \in (\zeta_{cn}, 0)$, since the amplitudes of the heavy fluid are $|\Phi_1| > |\Phi_2|$ for $\zeta \in (\zeta_{cn}, 0)$ with $|\Phi_1| = |\Phi_2|$ at $\zeta = \zeta_{cn}$, and the amplitudes of the light fluid are $|\tilde{\Phi}_1| > |\tilde{\Phi}_2|$ for $\zeta \in (\zeta_{cr}, 0)$ with $|\tilde{\Phi}_1| = |\tilde{\Phi}_2|$ at $\zeta = \zeta_{cr}$. The ratio is $\zeta_{cn}/\zeta_{cr} = 5/9$. For any Atwood number, the convergence limit bubble is less curved and has than larger velocity and smaller shear when compared to the critical bubble, Figure 5.

C. The Taylor bubble

The family has a solution with $\zeta = -(1/8)k$. We call this solution the Taylor bubble since its curvature is the same as in the solution which can be found in Davies & Taylor 1950, upon modifying the wavevector $k = 4\pi/\sqrt{3}\lambda$ to $k = 2\beta_1/\lambda$, where $\beta_1$ is the first zero of the Bessel function $J_1$ [4,11,12,37]. For the Taylor bubble, the curvature, the velocity, and the shear function are

$$\zeta_T = -\frac{1}{8}k, \quad v_T = \frac{1}{k\tau}\left(\frac{t}{\tau}\right)^{a/2} 2\sqrt{\frac{2A}{3+5A}}, \quad \Gamma_T = \frac{1}{\tau}\left(\frac{t}{\tau}\right)^{a/2} 3\sqrt{\frac{2A}{3+5A}} \quad (18)$$

Note that for the Taylor bubble the second Fourier amplitude of the heavy fluid is zero at $N = 1$ $(\Phi_2)_T = 0$. This property does not hold for $N > 1$. The ratios are $\zeta_T/\zeta_{cr} = 1/3$ and $\zeta_T/\zeta_{cn} = 3/5$.

For any Atwood number, the Taylor bubble is less curved and has larger velocity and smaller shear when compared to the convergent limit bubble and the critical bubble, Figure 5.

D. The Layzer-drag bubble

The bubble family has a solution with velocity $v = (k\tau)^{-1}(t/\tau)^{a/2}\sqrt{2A/(1+A)}$. We call the solution the Layzer-drag bubble since the drag model applies this velocity re-scaling with the Atwood number to the single-mode Layzer's first-order approximation at $A = 1$ [33,39]. Experiments and simulations tend to compare well with this re-scaling [33,39,47]. For the Layzer-drag bubble, the curvature, the velocity, and the shear function are

$$\zeta_L = \zeta_L(k, A), \quad v_L = \frac{1}{k\tau}\left(\frac{t}{\tau}\right)^{a/2}\sqrt{\frac{2A}{1+A}}, \quad \Gamma_L = \frac{1}{\tau}\left(\frac{t}{\tau}\right)^{a/2} 12\left(9 - 64\left(\frac{\zeta_L}{k}\right)^2\right)^{-1}\sqrt{\frac{2A}{1+A}} \quad (19)$$

The dependence of the Layzer-drag bubble curvature on the Atwood number is cumbersome, Figure 5. For fluids with very different densities $A \to 1^-$, the solution for the Layzer-drag bubble is:

$$A \to 1^-, \quad \zeta_L \approx -\frac{k}{8}, \quad v_L \approx \frac{1}{k\tau}\left(\frac{t}{\tau}\right)^{a/2}, \quad \Gamma_L \approx \frac{1}{\tau}\left(\frac{t}{\tau}\right)^{a/2}\frac{3}{2}$$

For fluid with very similar densities $A \to 0^+$, the solution for the Layzer-drag bubble is:

$$A \to 0^+, \quad \zeta_L \approx -\frac{k}{8}\sqrt{9 - 4\sqrt{3}}, \quad v_L \approx \frac{1}{k\tau}\left(\frac{t}{\tau}\right)^{a/2}\sqrt{2A}, \quad \Gamma_L \approx \frac{1}{\tau}\left(\frac{t}{\tau}\right)^{a/2}\sqrt{3}\sqrt{2A}$$

The Layzer-drag bubble is more curved for fluids with similar densities $A \to 0^+$ when compared to fluids with very different densities $A \to 1^-$.

For fluids with very different densities, $A \to 1^-$, the values of the curvature, velocity and shear of the Layzer-drag bubble are the same as the values of the corresponding quantities of the Taylor bubble. For fluids with a finite density ratio, $A \in (0,1)$ the curvatures relate as $\zeta_L/\zeta_{cr} \in \left(\sqrt{9 - 4\sqrt{3}}/3, 1/3\right)$ and $\zeta_L/\zeta_T \in \left(\sqrt{9 - 4\sqrt{3}}, 1\right)$. For $0 < A < 1$ the Layzer-drag bubble is more curved and has smaller velocity and larger shear when compared to the Taylor bubble, Figure 5.

E. The Atwood bubble

We call the fastest solution in the bubble family as the 'Atwood bubble' to emphasize its complex dependence on the Atwood number, $\zeta_A = \zeta_{max}$, $v_A = v_{max}$. The Atwood bubble solution has a remarkable invariant property [11,12,22,37,38]:

$$v_A \tau k \left(\frac{t}{\tau}\right)^{-a/2}\left(8\frac{|\zeta_A|}{k}\right)^{-3/2} = 1 \quad (20)$$

For fluids with very different densities $A \to 1^-$, the curvature, the velocity and the shear function of the Atwood bubble are:

$$A \to 1^-, \quad \zeta_A \approx -\frac{k}{8}, \quad v_A \approx \frac{1}{k\tau}\left(\frac{t}{\tau}\right)^{a/2}, \quad \Gamma_A \approx \frac{1}{\tau}\left(\frac{t}{\tau}\right)^{a/2}\frac{3}{2}$$

For fluid with very similar densities $A \to 0^+$ the curvature, the velocity and the shear function of the Atwood bubble are:

$$A \to 0^+, \quad \zeta_A \approx -\frac{3k}{16}A^{1/3}, \quad v_A \approx \frac{1}{k\tau}\left(\frac{t}{\tau}\right)^{a/2}(3/2)^{3/2}\sqrt{A}, \quad \Gamma_A \approx \frac{1}{\tau}\left(\frac{t}{\tau}\right)^{a/2}\sqrt{6A}$$

For fluids with very different densities, $A \to 1^-$, the values of the curvature, velocity and shear of the Atwood bubble are the same as the values of the corresponding quantities of the Taylor and the Layzer-drag bubbles. For fluids with a finite density ratio, the curvatures relate as $\zeta_A/\zeta_{cr} \in (0,1/3)$, $\zeta_A/\zeta_L \in (0,1)$ and $\zeta_A/\zeta_T \in (0,1)$. For $0 < A < 1$, the Atwood bubble is less curved and has larger velocity and smaller shear when compared to the Layzer-drag and the Taylor bubble, Figure 5.

F. The flat bubble

For the flat bubble with curvature $\zeta = \zeta_f$, the solution is:

$$\zeta_f = 0, \quad v_f = 0, \quad \Gamma_f = 0 \qquad (21)$$

The curvature, the velocity and the shear function of this bubble equal to zero. The ratio is $\zeta_f/\zeta_{cr} = 0$. Note that while the velocities of the flat and the critical bubbles are both zero, the shear of the flat bubble is zero, whereas the shear of the critical bubble has the maximum value.

**Sub-section 3.3.3 –Stability of solutions**

To analyze asymptotic stability of the solutions in the family, we slightly perturb $\zeta \to \zeta + \Delta\zeta$, $M \to M + \Delta M$ and $\tilde{M} \to \tilde{M} + \Delta\tilde{M}$ with $|\Delta\zeta/\zeta| << 1$, $|\Delta M/M| << 1$ and $|\Delta\tilde{M}/\tilde{M}| << 1$, such that $\Delta\zeta/\zeta \sim \psi(t/\tau)$, $\Delta M/M \sim \psi(t/\tau)$, and $\Delta\tilde{M}/\tilde{M} \sim \psi(t/\tau)$, where $\psi$ is some function of time $(t/\tau)$, $|\psi| << 1$, to be defined. Upon substituting these expressions to the governing equations Eqs.(7) and by linearizing the equations for small perturbations, we find that at $N = 1$

$$\psi = exp\left(\frac{\chi}{s}\left(\frac{t}{\tau}\right)^{s}\right), \quad s = \frac{a+2}{a} \qquad (22)$$

where $\chi$ is the function of the Atwood number and the bubble curvature. The real part of the function is negative $\mathrm{Re}[\chi]<0$ for stable solutions, and is positive $\mathrm{Re}[\chi]>0$ for unstable solutions [12,37,38].

For $a>-2$ the function $\chi$ does not depend on the acceleration parameters. The explicit expression for this function $\chi$ is cumbersome. The detailed investigation of the dependence of $\chi$ on the Atwood number, the bubble curvature, and the acceleration parameters suggests that for a finite Atwood number the flattened bubbles are unstable with $\mathrm{Re}[\chi]>0$, whereas the curved bubbles are stable with $\mathrm{Re}[\chi]<0$, including the Atwood bubble, the Layzer-drag bubble, the Taylor's bubble, and the convergence limit bubble [22]. The $N>1$ analysis is similar to [37,38], and we address it to the future.

**Sub-section 3.4 – Dynamics of nonlinear spikes**

**Sub-section 3.4.1 – Nonlinear solutions**

For nonlinear spikes, we consider solutions with $v(\tau k)<0, (\zeta/k)>0$ in the family in Eqs.(15), Table 2. In the family of asymptotic solutions with positive curvatures $\zeta/k>0$ in Eqs.(15), velocity values are negative, $v(\tau k)<0$, and the shear function values are negative, $\Gamma\tau<0$. Nonlinear spikes move down and are concave up; vortical structures, which may appear in a vicinity of a spike due to the interfacial shear, 'rotate' from the light to the heavy fluid, in agreement with experiments [14-18].

The solutions with $(\zeta/k)>0$, $v(\tau k)<0$ and $\Gamma\tau<0$ have the following properties. Domain of the velocity function $v$ is $\zeta\in(\hat{\zeta}_{max},\hat{\zeta}_{cr})$ with $\hat{\zeta}_{max}\in[0,\hat{\zeta}_{cr}]$. For $\zeta\in(\hat{\zeta}_{max},\hat{\zeta}_{cr})$, the range of the function $v$ is $v\in(\hat{v}_{max},0)$ with $v\to 0$ for $\zeta\to\hat{\zeta}_{cr}$ and with $v\to\hat{v}_{max}$ for $\zeta\to\hat{\zeta}_{max}$, Figure 3. The critical curvature value is $\hat{\zeta}_{cr}=(3/8)k$. The curvature value $\hat{\zeta}_{max}$ bounds the interval of values $\zeta$, in which the function $v$ is real and non-positive. The value $\hat{\zeta}_{max}$ is the pole of the velocity function as well as the amplitudes and the shear function, and it is $\hat{\zeta}_{max}/k=3\left(1-\sqrt{1-A^2}\right)/(8A)$. It corresponds to the spike with the largest value of the velocity magnitude $\hat{v}_{max}$, with $\hat{v}_{max}\to-\infty$. The shear function is $\Gamma\in\left(-|\hat{\Gamma}_{max}|,-|\hat{\Gamma}_{min}|\right)$ for $\zeta\in(\hat{\zeta}_{max},\hat{\zeta}_{cr})$, achieving the value with the minimum magnitude $\Gamma\to\hat{\Gamma}_{min}$, $\hat{\Gamma}_{min}<0$, for $\zeta\to\hat{\zeta}_{cr}$ and achieving the value with the maximum magnitude of the shear $\Gamma\to\hat{\Gamma}_{max}$, $\hat{\Gamma}_{max}<0$, approaching negative infinity, $\hat{\Gamma}_{max}\to-\infty$, for $\zeta\to\hat{\zeta}_{max}$, Figure 3.

For nonlinear RT spikes, Figure 6a, Figure 6b, Figure 6c illustrate the dependence of the re-scaled velocity $v/\overline{v}_{max}$ and the re-scaled shear function $\Gamma/\overline{\Gamma}_{max}$ on the re-scaled curvature $\zeta/\hat{\zeta}_{cr}$, as well as the dependence of the shear $\Gamma/\overline{\Gamma}_{max}$ on the velocity $v/\overline{v}_{max}$ for some Atwood numbers $A$ respectively. Here the scaling values are regularized as $\overline{v}_{max} = \hat{v}_{max}\left(\left(\zeta - \hat{\zeta}_{max}\right)/k\right)^{1/2}$ and $\overline{\Gamma}_{max} = \hat{\Gamma}_{max}\left(\left(\zeta - \hat{\zeta}_{max}\right)/k\right)^{1/2}$. The universality of nonlinear dynamics of RT spikes and the link between the interface dynamics, interface morphology and the interfacial shear are clearly seen from these plots. Figure 6d further illustrates the convergence of solutions for nonlinear of RT spikes with $\zeta = \hat{\zeta}_{max}$ by showing that for a broad range of values of the Atwood numbers $A$ the lowest-order amplitudes $|\Phi_1|, |\tilde{\Phi}_1|$ are dominant even at the (singular) point $\zeta = \hat{\zeta}_{max}$.

According to our results, late-time dynamics of nonlinear RT spikes may exhibit singular behavior, in agreement with experiments, Figure 3, Figure 6 [11-18].

**Sub-section 3.4.2 – Special solutions**

The family of asymptotic solutions for nonlinear RT spikes with $\zeta/k > 0$ and $v(\tau k) < 0$ has some special solutions. We call these solutions the critical spike, the convergence limit spike, the Taylor spike, the Layzer-drag spike, the Atwood spike, and the flat spike. Properties of these solutions are illustrated by Figure 7.

A. The critical spike

For the critical spike with curvature $\zeta = \hat{\zeta}_{cr}$, the solution is

$$\hat{\zeta}_{cr} = \frac{3}{8}k, \quad \hat{v}_{cr} = 0, \quad \hat{\Gamma}_{cr} = -\frac{1}{\tau}\left(\frac{t}{\tau}\right)^{a/2}\sqrt{\frac{6A}{1-A}}, \quad \hat{\Gamma}_{cr} = -\left|\hat{\Gamma}_{min}\right|, \quad \hat{\Gamma}_{min} < 0 \quad (23)$$

For the critical spike, the velocity is zero for any density ratio, and the shear function has a non-positive value, which is zero for $A \to 0$ and is finite for $0 < A < 1$ and which is singular approaching negative infinity for $A \to 1$, Figure 7.

The magnitudes of the curvature and the velocity of the critical spike are the same as those for the critical bubble, with $\hat{\zeta}_{cr} = -\zeta_{cr}$ and $\hat{v}_{cr} = v_{cr} = 0$. The magnitude of the shear of the critical spike differs from that of the critical bubble: The shear's magnitude of the critical spike approaches infinity as $A \to 1^-$, whereas the shear's magnitude of the critical bubble is finite for any $A$, Figure 5, Figure 7.

B. The convergence limit spike

At $N=1$, the amplitudes of the heavy fluid relate as $|\Phi_1|>|\Phi_2|$ for $\zeta\in(\hat{\zeta}_{max},\hat{\zeta}_{cr})$, with $|\Phi_1|=|\Phi_2|=0$ at $\hat{\zeta}_{cr}=(3/8)k$. The light fluid amplitudes relate as $|\tilde{\Phi}_1|>|\tilde{\Phi}_2|$ for $\zeta\in(\hat{\zeta}_{max},\hat{\zeta}_{cn})$, with $|\tilde{\Phi}_1|=|\tilde{\Phi}_2|$ at $\hat{\zeta}_{cn}=(5/24)k$ (sub-script stands for convergence). This defines the solution for the convergence limit spike.

For the convergence limit spike with curvature $\zeta=\hat{\zeta}_{cn}$, the solution is

$$\hat{\zeta}_{cn}=\frac{5}{24}k,\quad \hat{v}_{cn}=-\frac{1}{k\tau}\left(\frac{t}{\tau}\right)^{a/2}\frac{14}{3}\sqrt{\frac{10A}{3(45-53A)}},\quad \hat{\Gamma}_{cn}=-\frac{1}{\tau}\left(\frac{t}{\tau}\right)^{a/2}9\sqrt{\frac{10A}{3(45-53A)}} \qquad (24)$$

At $N=1$, the convergence holds for $\zeta\in(\hat{\zeta}_{max},\hat{\zeta}_{cn})$, since the amplitudes are $|\tilde{\Phi}_1|>|\tilde{\Phi}_2|$ and $|\Phi_1|>|\Phi_2|$ for $\zeta\in(\hat{\zeta}_{max},\hat{\zeta}_{cn})$ with $|\tilde{\Phi}_1|=|\tilde{\Phi}_2|$ at $\zeta=\hat{\zeta}_{cn}$, whereas the amplitudes $|\Phi_1|>|\Phi_2|$ for $\zeta\in(\hat{\zeta}_{max},\hat{\zeta}_{cr})$, Figure 6, Figure 7.

For the convergence limit spike, the velocity and the shear function have non-positive values, which are zero for $A\to 0^+$ and are finite for $0<A<45/53$ and which become singular approaching negative infinity for $A\to 45/53\approx 0.85$. The ratio is $\hat{\zeta}_{cn}/\hat{\zeta}_{cr}=5/9$, and the convergence limit spike is less curved when compared to the critical spike, Figure 7.

The magnitude of the curvature of the convergence limit spike is the same as that for the convergence limit bubble, with $\hat{\zeta}_{cn}=-\zeta_{cn}$. The magnitudes of the velocity and shear of the convergence limit spike differ from their respective values of the convergence limit bubble: The magnitudes of the velocity and shear of the spike approaches infinity for $A\to 45/53$, whereas the magnitudes of the velocity and shear of the bubble are finite for any $A$. Note that the convergence limit spike is defined by the condition $|\tilde{\Phi}_1|=|\tilde{\Phi}_2|$ for the amplitudes of the light fluid, whereas the convergence limit bubble is defined by the condition $|\Phi_1|=|\Phi_2|$ for the amplitudes of the heavy fluid, Figure 3, Figure 5, Figure 7.

C. The Taylor spike

The family has a solution with $\zeta=(1/8)k$. We call this solution the Taylor spike, by analogy with the Taylor bubble [11,12,37].

For the Taylor spike with curvature $\zeta=\hat{\zeta}_T$, the solution is

$$\hat{\zeta}_T = \frac{1}{8}k, \quad \hat{v}_T = -\frac{1}{k\tau}\left(\frac{t}{\tau}\right)^{a/2} 2\sqrt{\frac{2A}{3-5A}}0, \quad \hat{\Gamma}_T = -\frac{1}{\tau}\left(\frac{t}{\tau}\right)^{a/2} 3\sqrt{\frac{2A}{3-5A}} \quad (25)$$

For this special solution, the second Fourier amplitude of the light fluid is zero $\left(\tilde{\Phi}_2\right)_T = 0$ at $N = 1$. This property does not hold for $N > 1$, Figure 7.

For the Taylor spike, the velocity and the shear function have non-positive values, which are zero for $A \to 0$ and are finite for $0 < A < 3/5$ and which approach negative infinity and become singular for $A \to 3/5 = 0.6$. Since the ratios are $\hat{\zeta}_T / \hat{\zeta}_{cr} = 1/3$ and $\hat{\zeta}_T / \hat{\zeta}_{cn} = 3/5$, the Taylor spike is less curved when compared to the critical spike and the convergence limit spike, Figure 7.

The magnitude of the curvature of the Taylor spike is the same as that for the Taylor bubble, with $\hat{\zeta}_T = -\zeta_T$. The magnitudes of the velocity and shear of the Taylor spike differ from their respective values of the Taylor bubble: The velocity and shear magnitudes for the spike approach infinity for $A \to 3/5$; the velocity and shear magnitudes for the bubble are finite for any $A$, Figure 5, Figure 7.

D. The Layzer-drag spike

RT family of spikes has a solution with velocity $v = -(k\tau)^{-1}(t/\tau)^{a/2}\sqrt{2A/(1-A)}$. We call this solution the Layzer-drag spike since the drag model applies such re-scaling with the Atwood number for the velocity of the nonlinear spike to the Layzer first order approximation of the velocity of the nonlinear bubble at $A = 1$ [33,39]. Experiments and simulations tend to compare well with this re-scaling [39,47]. For the Layzer-drag spike, the curvature, the velocity, and the shear function are

$$\hat{\zeta}_L = \hat{\zeta}_L(k, A), \quad \hat{v}_L = -\frac{1}{k\tau}\left(\frac{t}{\tau}\right)^{a/2}\sqrt{\frac{2A}{1-A}}, \quad \hat{\Gamma}_L = -\frac{1}{\tau}\left(\frac{t}{\tau}\right)^{a/2} 12\left(9 - 64\left(\frac{\hat{\zeta}_L}{k}\right)^2\right)^{-1}\sqrt{\frac{2A}{1-A}} \quad (26)$$

The dependence of the Layzer-drag spike curvature on the Atwood number is cumbersome. The velocity and the shear function of the Layzer-drag spike have non-positive values, that are zero for $A \to 0$ and are finite for $0 < A < 1$ and that become singular approaching negative infinity for $A \to 1$.

For fluids with very different densities $A \to 1^-$, the solution for the Layzer-drag spike is:

$$A \to 1^-, \quad \hat{\zeta}_L \to \frac{3}{8}k, \quad \hat{v}_L \approx -\frac{1}{k\tau}\left(\frac{t}{\tau}\right)^{a/2}\sqrt{\frac{2}{1-A}}, \quad \hat{\Gamma}_L \approx -\frac{1}{\tau}\left(\frac{t}{\tau}\right)^{a/2}\frac{1}{4}\left(\frac{3}{8} - \frac{\zeta_L}{k}\right)^{-1}\sqrt{\frac{2}{1-A}}$$

The velocity and the shear of the Layzer-drag spike are singular for $A \to 1^-$, with the shear is even more singular than the velocity.

For fluid with very similar densities $A \to 0^+$, the solution for the Layzer-drag spike is:

$$A \to 0^+, \quad \hat{\zeta}_L \approx \frac{k}{8}\sqrt{9-4\sqrt{3}}, \quad v_L \approx -\frac{1}{k\tau}\left(\frac{t}{\tau}\right)^{a/2}\sqrt{2A}, \quad \hat{\Gamma}_L \approx -\frac{1}{\tau}\left(\frac{t}{\tau}\right)^{a/2}\sqrt{3}\sqrt{2A}$$

The Layzer-drag spike is more curved for fluids with very different densities $A \to 1^-$ than for fluids with very similar densities $A \to 0^+$. For a two fluid system with $A \in (0,1)$ the ratios are $\hat{\zeta}_L / \hat{\zeta}_{cr} \in \left(\sqrt{9-4\sqrt{3}}/3, 1\right)$ and $\hat{\zeta}_L / \hat{\zeta}_T \in \left(\sqrt{9-4\sqrt{3}}, 3\right)$, and the Layzer-drag spike is less curved than the critical spike and is more curved than the Taylor spike, Figure 7. The Layzer-drag spike has the smaller magnitude of the velocity and the smaller magnitude of the shear when compared to those of the Taylor spike. The curvatures of the Layzer-drag spike and the convergence limit spike equal one another $\hat{\zeta}_L / \hat{\zeta}_{cn} = 1$ at $A = 235/451 \approx 0.52$, and for larger values of the Atwood numbers the Layzer-drag spike solutions are not convergent, Figure 7.

For fluids with very similar densities, $A \to 0$, the magnitude of the curvature of the Layzer-drag spike is the same as that for the Layzer-drag bubble, with $\hat{\zeta}_L = -\zeta_L$. The magnitudes of the velocity and shear of the Layzer-drag spike differ from those of the Layzer-drag bubble: The velocity and shear magnitudes of the spike approach infinity for $A \to 1$, whereas the velocity and the shear magnitudes are finite for any $A$, Figure 5, Figure 7.

E. The Atwood spike

The fastest spike in the family of solutions is $\left(\hat{\zeta}_{max}, \hat{v}_{max}\right)$. We call this solution the 'Atwood spike' due to its complex dependence on the Atwood number, $\hat{\zeta}_A = \hat{\zeta}_{max}$, $\hat{v}_A = \hat{v}_{max}$. This solution is

$$\hat{\zeta}_A = \kappa(A)\,k, \quad \hat{v}_A \approx -\frac{\varphi(A)}{k\tau}\left(\frac{t}{\tau}\right)^{a/2}\left(\frac{k}{\zeta-\hat{\zeta}_A}\right)^{1/2}, \quad \hat{\Gamma}_A \approx -\frac{\gamma(A)}{\tau}\left(\frac{t}{\tau}\right)^{a/2}\left(\frac{k}{\zeta-\hat{\zeta}_A}\right)^{1/2} \qquad (27)$$

For the Atwood spike, the curvature is regular, $\zeta \to \hat{\zeta}_A$, whereas the velocity and the shear function are singular, with $\hat{v}_A \to -\infty, \hat{\Gamma}_A \to -\infty$ for any Atwood number. The functions $\kappa, \varphi, \gamma$ for the curvature, the velocity and the shear are respectively

$$\kappa(A) = \frac{3\left(1-\sqrt{1-A^2}\right)}{8A}, \quad \varphi(A) = \frac{9}{4}\frac{\left(1-A^2\right)^{1/4}\left(1-\sqrt{1-A^2}\right)^{3/2}}{A^2}, \quad \gamma(A) = \frac{3}{2}\frac{\left(1-\sqrt{1-A^2}\right)^{1/2}}{\left(1-A^2\right)^{1/4}} \qquad (28)$$

For fluids with very different densities $A \to 1^-$, these functions behave as

$$A \to 1^-, \quad \kappa \to \frac{3}{8}\left(1-(2(1-A))^{1/2}\right), \quad \varphi \to \frac{9}{4}(2(1-A))^{1/4}, \quad \gamma \to \frac{3}{2}(2(1-A))^{-1/4}$$

Note the singular character of the function $\gamma$ and hence $\hat{\Gamma}_A$ for $A \to 1^-$. For fluids with very similar densities $A \to 0^+$, these functions behave as

$$A \to 0^+, \quad \kappa \to \frac{3}{16}A, \quad \varphi \to \frac{9}{8\sqrt{2}}A, \quad \gamma \to \frac{3}{2\sqrt{2}}A$$

The Atwood spike is less curved than the Layzer-drag spike for $A \in [0,1)$ and has the same curvature as the Layzer-drag spike $\hat{\zeta}_A / \hat{\zeta}_L = 1$ at $A = 1$. The Atwood spike attains the same curvature as the critical spike $\hat{\zeta}_A / \hat{\zeta}_{cr} = 1$ at $A = 1$, the same curvature as the Taylor's spike $\hat{\zeta}_A / \hat{\zeta}_T = 1$ at $A = 3/5 \approx 0.6$, and the same curvature as the convergence limit spike $\hat{\zeta}_A / \hat{\zeta}_{cn} = 1$ at $A = 45/53 \approx 0.85$, Figure 7. For $A > 45/53$ the Atwood spike solutions may not converge.

The properties of the curvature, velocity and shear Atwood spike differ dramatically from those of the Atwood bubble: For any density ratio $A$, the magnitudes of the velocity and shear of the Atwood spike are singular whereas the magnitudes of the velocity and shear of the Atwood bubble are regular and finite. While the magnitudes of the curvature of the Atwood spike and the Atwood bubble are both finite, the Atwood spike is more curved (less curved) when compared to the Atwood bubble for fluids with very different (very similar) densities $A \to 1^-$ ($A \to 0^+$), Figure 5, Figure 7.

F. The flat spike

For the flat spikes with curvature $\zeta = \hat{\zeta}_f$, the solution is:

$$\hat{\zeta}_f \equiv 0, \quad \hat{v}_f \equiv 0, \quad \hat{\Gamma}_f \equiv 0 \quad (29)$$

The curvature, the velocity and the shear function of this spike equal to zero. The ratio is $\hat{\zeta}_f / \hat{\zeta}_{cr} = 0$. Note that while the value $\hat{\zeta}_A \to \hat{\zeta}_f$ for $A \to 0^+$, the flat spike remains an isolated solution, in contrast to the flat bubble.

**Sub-section 3.4.3 – Stability of solutions**

Stability of asymptotic solutions in the family of spikes can be studied similarly to the case of the family of bubbles, Eqs.(22). The stability properties of the spikes differ significantly from those of the bubbles. Particularly, in the family of spikes, for a given Atwood number, there is a narrow interval of stable solutions with finite curvatures, whereas fast solutions with $\zeta \to \hat{\zeta}_A$ are unstable. This suggests that nonlinear spikes can be in transient and move faster, with $|v\,\tau\,k| \sim (t/\tau)^{\delta}$, $\delta > a/2$, than it is prescribed by the nonlinear asymptotic dynamics, with $|v\,\tau\,k| \sim (t/\tau)^{a/2}$.

**Sub-section 3.5 – Properties of Rayleigh-Taylor dynamics**

**Sub-section 3.5.1 – Interplay of acceleration, morphology, and shear**

By analyzing properties of the linear and nonlinear dynamics Eqs.(7-29) for $a > -2$, we find that in the linear regime RT dynamics is the acceleration driven: the growth-rate of the interface perturbations is defined by the acceleration, whereas the initial morphology of the interface defines the positions of bubbles and spikes Eqs.(9,10) Table 1 [22,23]. The nonlinear dynamics is more complex. From the properties of the solutions' family Eqs.(15), including the special solutions for RT bubbles Eqs.(16-21) and for RT spikes Eqs.(23-29), we find that for incompressible ideal fluids the asymptotic nonlinear RT dynamics is influenced by the interplay of the acceleration, the interface morphology and the interfacial shear, Table 2, Figure 3 - Figure 7.

A. RT bubbles

From the properties of the solutions describing RT bubbles it follows that the curved bubbles move faster than the flattened bubbles due to the larger effective acceleration, Eqs.(15,16-21), Figure 4. The larger curvature leads to the larger interfacial shear. The larger shear, in turn, stronger decelerates the bubble. For a well curved bubble, the shear is large, the deceleration is strong, and the bubble velocity decreases, Figure 4. The shear is 1-1 function on the curvature, and, for the critical bubble, the very large shear can alone maintain the pressure at the interface, Figure 4, Figure 5.

For any Atwood number, the bubble velocity as the function of the interfacial shear is very nearly linear for small shear values, then it achieves its maximum value corresponding to the Atwood bubble, and then it sharply drops to zero for large shear values, Eqs.(20), Figure 4. In a vicinity of the fastest moving Atwood bubble solution, the velocity has a very steep dependence on the interfacial shear. This suggests the need in highly accurate methods of numerical modeling and experimental diagnostics for accurate capturing of RT dynamics [13-21].

B. RT spikes

Similarly to RT bubbles, RT spikes move slower for larger magnitudes of their curvatures, Eqs.(15,23-29), Figure 6. For fluids with a finite density ratio, the shear of the strongly curved critical spike can alone maintain the pressure at the interface reducing the spike velocity to zero value, Figure 6. In contrast to RT bubbles, RT spikes are singular, Figure 3, Figure 6. The velocity and the shear values of the convergence limit and the Taylor spikes become singular approaching negative infinity at some Atwood numbers, Figure 7. Those of the Layzer-drag spike approach negative infinity for the Atwood number $A \to 1^-$, with the shear being more singular than the velocity, Figure 7.

The fastest moving Atwood spike is singular for any Atwood number, Eqs.(27,28), Figure 7. For the Atwood spike, the velocity and shear both approach infinite values in case of fluids with a finite density ratio $0 < A < 1$, and the shear is even more singular than the velocity in case of fluids with highly contrasting densities $A \to 1^-$. The singular character and the instability of the Atwood spike solution suggest that RT spike, upon achieving a finite curvature value, can become unstable and move faster than it is prescribed by the nonlinear asymptotic dynamics Eqs.(14).

**Sub-section 3.5.2 – Single-scale and multi-scale dynamics**

By analyzing properties of the early-time linear dynamics Eqs.(10), we find that for ideal incompressible fluids and for the interface perturbations with very small initial amplitude RT, dynamics is singe-scale and is defined by the horizontal length scale, which is the spatial period (the wavelength) of the interface perturbation, Figure 2.

The nonlinear dynamics is more complex. From the properties of the solutions' family in Eqs.(15), we find that for ideal incompressible fluids the asymptotic nonlinear RT dynamics is essentially multi-scale and is defined by the two macroscopic scales – the horizontal length scale (the wavelength) and the vertical length scale (the amplitude). This can be understood by viewing RT coherent structure as a standing wave, whose nonlinear dynamics is influenced by its wavelength and its growing amplitude.

A. RT bubbles

In the family of the asymptotic solutions describing nonlinear RT bubbles, we define the fastest stable solution as the physically significant solution, $\zeta \sim \zeta_A, v \sim v_A, \Gamma \sim \Gamma_A$. The Atwood bubble has the invariant property, $v_A \tau k \left(t/\tau\right)^{-a/2} \left(8\left|\zeta_A\right|/k\right)^{-3/2} = 1$, Eqs.(20). The left-hand side of the expression is a function of time, and of the vertical and the horizontal length scales and their derivatives, whereas the right-hand side is unity. This invariance implies that the two macroscopic scales - the wavelength $\lambda$ and amplitude $z_0$ - contribute to the multi-scale dynamics of nonlinear RT bubbles [11,12,22,37,38].

B. RT spikes

The fastest Atwood spike has no invariant property similar to that of the fastest Atwood bubble. The singular character of the dynamics of the (unstable) fastest Atwood spike with $\left|\hat{v}_A\right|\left(Gt^a/k\right)^{-1/2} \to \infty$ suggests that the nonlinear RT spike can move faster than the nonlinear asymptotic dynamics prescribes, Eqs.(27,28). As such, it is influenced even more by the vertical scale (the amplitude) when compared to the nonlinear RT bubbles. The dynamics of the nonlinear RT spike is thus multi-scale, with the two macroscopic length scale contributing - the vertical length scale (amplitude) and the horizontal length scale (wavelength). To our knowledge, this is the first report of the multi-scale character of nonlinear dynamics of RT spikes based on the properties of nonlinear asymptotic solutions for the boundary value problems Eqs.(1-4).

**Sub-section 3.5.3 – Transition to the self-similar mixing**

The linear and nonlinear RT dynamics are scale-dependent. As time evolves, the flow transits to the scale-invariant regime, in which the vertical scale grows self-similarly with time [3,4,11-22,39-47]. Group theory approach studies properties of self-similar RT dynamics within the frame of momentum model [11,14,22,42,55]. This model has the same symmetries and scaling transformations as the governing equations and yields asymptotic solutions for the scale-dependent and the self-similar dynamics, up to constant.

The important outcome of the momentum model is that RT dynamics can transit from the scale-dependent nonlinear regime to the scale-invariant self-similar mixing regime when the vertical scale (the amplitude) is the dominant scale, in excellent agreement with high Reynolds number experiments [11,14,42]. This mechanism compliments the traditional merger mechanism, in which the transition to the self-similar mixing is believed to occur via the growth of horizontal scales [39,40,41]. Moreover, for RT coherent structures with hexagonal symmetry the growth of horizontal scales may not be feasible, since it may lead to anisotropy of the highly isotropic hexagonal pattern [56]. Ref.[56] provides the detailed classification of the formation of patterns in RT flows and possible discrete transitions with the increase of the wavelength(s) of RT coherent structures under the influence of modulations. These results excellently agree with existing experiments [13-18].

We address to the future the identification of a direct link between the analytical results in the present paper and momentum model. Here we analyze how the multi-scale character of the nonlinear dynamics is associated with and may lead to the acceleration of RT bubbles and RT spikes.

A. RT bubbles

Recall that in laboratory frame of reference the interface is described near the tip of the bubble as $Z = z_0 + \zeta\left(x^2 + y^2\right)$ with the velocity $v = \dot{z}_0$. For the fastest Atwood bubble the curvature is $\zeta \sim \zeta_A$ and the velocity is $v \sim v_A$, and their values are $|\zeta_A| \sim k \sim \lambda^{-1}$ and $|v_A| \sim (k\tau)^{-1}(t/\tau)^{a/2}$, as found by the asymptotic balance $v^2 \sim g\,|\zeta| \sim g/\lambda$, leading to $|v| \sim \sqrt{G\lambda t^a}$ and $|z_0| \sim t\sqrt{G\lambda t^a}$ Eqs.(14,15). The invariant property of the Atwood bubble Eqs.(20) suggests that the amplitude $z_0$ and the spatial period $\lambda$ contribute independently to the multi-scale dynamics of the nonlinear bubble. Hence the amplitude $z_0$ can be a dominant scale defining the rate of loss of specific momentum (i.e., the drag force per unit mass) as $\sim v^2/|z_0|$ [42,55]. The effective reduction of the drag force from $\sim v^2/\lambda$ to $\sim v^2/|z_0|$ can lead (for $|z_0| > \lambda$) to the bubble acceleration and transit the bubble from the scale-dependent nonlinear regime with $|v| \sim \sqrt{G\lambda t^a}$ and $|z_0| \sim t\sqrt{G\lambda t^a}$ to self-similar mixing with $|v| \sim Gt^{a+1}$ and $|z_0| \sim Gt^{a+2}$ [11,14,22, 42].

B. RT spikes

Similarly to RT bubbles, in laboratory frame of reference the interface is described near the tip of the spike as $Z = z_0 + \zeta\left(x^2 + y^2\right)$ with the velocity $v = \dot{z}_0$. For the Atwood spike the curvature is $\hat{\zeta} \sim \hat{\zeta}_A$ and the velocity is $v \sim \hat{v}_A$ with the finite value of the curvature $\left|\hat{\zeta}_A\right| \sim k \sim \lambda^{-1}$ and with the singular value of the velocity approaching infinity $|\hat{v}_A| \sim (k\tau)^{-1}(t/\tau)^{a/2}\left(k/\left(\zeta - \hat{\zeta}_A\right)\right)^{1/2} \to \infty$ Eqs.(27,28). This suggests that for the Atwood spike the amplitude $|z_0|$ can increase faster than it is prescribed by the asymptotic balance $v^2 \sim g\,|\zeta|/k^2 \sim g\lambda$ with $|v| \sim \sqrt{G\lambda t^a}$ and $|z_0| \sim t\sqrt{G\lambda t^a}$ in Eqs.(14,15). Thus the amplitude $z_0$ can be the dominant scale defining the rate of loss of specific momentum (i.e., the drag force per unit mass) as $\sim v^2/|z_0|$. The reduction of the rate of loss of specific momentum (the drag force per unit mass) from $\sim v^2/\lambda$ to $\sim v^2/|z_0|$ can lead (for $|z_0| >> \lambda$,) to the acceleration of the spike and transits the spike from the scale-dependent nonlinear dynamics to the self-similar mixing with $|z_0| \sim Gt^{a+2}$ and $|v| \sim Gt^{a+1}$ [11,14,22,42,55].

For fluids with very different densities, $A \to 1^-$, the Atwood spike is highly singular, suggesting that its dynamics can only be self-similar, in agreement with experiments [4,13-19]. To our knowledge, this is the first analysis of the mechanism of transition to self-similar mixing on the basis of properties of rigorous solutions for nonlinear dynamics of RT bubbles and RT spikes in Eqs.(1-4).

**Sub-section 3.5.4 – Interfacial dynamics**

By accurately accounting for the interplay of harmonics and by systematically connecting the interface velocity to the interfacial shear in a broad range of the acceleration parameters, our analysis finds that RT dynamics is essentially interfacial: In both the linear and nonlinear regimes, for both RT bubbles and RT spikes, the flow has intense motion of the fluids in a vicinity of the interface and has effectively no motion away from the interface. The velocity field is potential in the fluid bulk. Due to the presence of shear at the interface, shear-driven vortical structures develop at the interface. Figure 8 illustrate the velocity fields in the vicinity of the tips of RT bubble and RT spike.

This velocity pattern is in excellent agreement with experiments and simulations [13-21]. For realistic fluids, in order to accurately grasp the formation and the dynamics of vortical structures, one should also account for the influence of viscosity, compressibility, surface tension, mass flux across the interface, magnetic field, and other factors. We address these tasks to the future.

**Sub-section 3.5.5 – Effect of interfacial shear**

Our theory finds that the interfacial shear plays the important role in RT dynamics. The presence of interfacial shear is fully consistent with the conversation laws, including the continuity of tangential component of momentum at the interface, and, in case of zero interfacial mass flux, the discontinuity of tangential component of velocity at the interface, Eqs.(1-3). The presence of shear at the interface leads to appearance of the shear-driven vortical structures at the interface, Figure 8 [13-18]. It is a common wisdom that RT evolution is accompanied by shear-driven Kelvin-Helmoltz instability (KHI) developing due to the jump of tangential component of velocity at the interface [13-21]. Our results can be used to explain and predict the properties of KHI-driven vortical structures in RT flow.

First, according to our results, the jump of tangential velocity at the interface is zero at the tips of the bubble and the spike, and it increases away from these tips. Hence, the interface is stable and is free from vortical structures exactly at the tips of the bubble and the spike. KHI may develop away from the tips of the bubble or spike leading to the appearance of shear-driven vortical structures on the sides of evolving bubbles and spikes. This theoretical explanation of the stability of the tips of bubbles and spikes in RT flow agrees excellently with experiments [13-18].

Second, the shear function, which we establish in our work, can be used to estimate the growth-rate of KHI, as $\omega_{KHI} \sim |\Gamma|$ in the region of bubbles and $\hat{\omega}_{KHI} \sim |\hat{\Gamma}|$ in the region of spikes, and to find the dependence of $\omega_{KHI}, \hat{\omega}_{KHI}$ on the density ratio, interface morphology, flow symmetry, and acceleration parameters. According to our results, the shear function is regular in the region of bubbles and is singular in the region of spikes. Hence, one expects a more intense production of vortical structures for RT spikes when compared to RT bubbles, in excellent agreement with existing observations [13-21]. To our knowledge, this is the first theoretical explanation of the bubble/spike asymmetry of vortical structures in RT flow.

**Sub-section 3.5.6 – Formal properties of nonlinear solutions**

By directly linking the interface velocity, the flow fields, the interfacial shear function, and the interface morphology in RT nonlinear dynamics, our theory finds the physics interpretation and resolves the long-standing problem of multiplicity of solutions for nonlinear RTI [33-37]. According to our results, for RT dynamics with variable acceleration, there is a continuous family of asymptotic nonlinear solutions, with each solution having its own curvature, velocity, the Fourier amplitudes, the flow fields, and the interfacial shear [22]. The multiplicity of nonlinear asymptotic solutions for the boundary value problem is due to the presence of shear at the interface, and is the essential property of RT dynamics associated with the governing equations and the boundary conditions at the interface Eqs.(1-4) [11,22,35]. In the continuous family of nonlinear solutions, asymptotic solutions with negative curvature and positive velocity correspond to regular RT bubbles (which move up and concave down), whereas asymptotic solutions with positive curvature and negative velocity correspond to singular RT spikes (which move down and concave up), Figure 1, Figure 8. To our knowledge, our work is the first to report that asymptotic solutions describing dynamics of the nonlinear RT spikes and the nonlinear RT bubbles with variable acceleration form a continuous family.

**Sub-section 3.5.7 – Effect of the acceleration parameters**

According to our results, for a broad range of the acceleration parameters, $a > -2, G > 0$, the interface dynamics is described by tabular special functions, such as modified Bessel functions and power-law functions. The early-time linear dynamics is super-exponential (i.e. quicker than exponential) for $a > 0$, is exponential at $a = 0$, and is sub-exponential (i.e. slower than exponential) for $-2 < a < 0$, Eqs.(10), Table 1, Figure 2. For the late-time nonlinear dynamics the velocity near the regular point of the interface increases with time for $a > 0$, is steady at $a = 0$, and decrease with time for $a < 0$, Eqs.(15), Table 2, Figure 3. Furthermore, except for the difference in the time-dependence, nonlinear dynamics depends only on the interface morphology and the flow symmetry.

These properties enable a comparative study of RT dynamics for various acceleration parameters, $a > -2, G > 0$. Particularly, by analyzing properties of fast RT dynamics for large exponents $a > 0$, one can deduce properties of slow RT dynamics for small exponents $-2 < a < 0$. Such analysis is especially important in studies of RTI in high energy density plasmas (for instance, in supernovae and fusion), where RTI is driven by an explosion or an implosion with acceleration exponents $-2 < a < -1$ defined by blast waves [22,50-54]. Our results thus provide new extensive theory benchmarks for the future.

**Sub-section 3.5.8 – Comparisons with existing observations**

Significant efforts are undertaken in RT community to better understand the scale-dependent RT dynamics and the self-similar RT mixing. See for details edited research books [1,2] and experimental and numerical papers [11-25,45-47,55,-60]. As discussed earlier in Sections 3.1.1-3.5.7, our theory is in excellent qualitative agreement with the existing observations [11-25,45-47,55,-60], and provides new extensive quantitative benchmarks for experiments and simulations. To date, to the best of the authors' knowledge, accurate experiments and simulations are free from systematic investigations of RT dynamics driven by variable acceleration for three-dimensional flow with hexagonal symmetry [13-21].

In this work we consider RT dynamics with variable acceleration for three-dimensional flow with hexagonal symmetry, since it is important to study from the physics perspectives and since it has a number of advantages when compared to other space groups. For instance, hexagonal group is the most isotropic, when compared to other space groups; for structures with hexagonal symmetry, transitions to super-structures with larger periods lead to a loss of isotropy and may thus be infeasible [10-12]. At the same time, structures with hexagonal symmetry may be challenging to accurately implement in experiments and simulations. To the best of the authors' knowledge, the existing experiments and numerical simulations are absent for such setup [13-21]. Our theory results provide new benchmarks for the future research.

For instance, the agreements and departures of our scale-dependent solutions for the linear and nonlinear dynamics of RT bubbles and RT spikes from those in experiments and numerical simulations can serve to better understand RT dynamics in realistic fluids. Note that our theory focuses on the large-scale dynamics and presumes that interfacial vortical structures are small. This assumption is applicable for fluids very different densities $A \to 1^{-}$ and with finite density ratios $0 << A < 1$. For fluids with very similar densities $A \to 0^{+}$ other approaches can be employed [11,12,22].

Our theory finds the effect of the interplay of the acceleration, the interface morphology and the interfacial shear on the scale-dependent dynamics of RTI with variable acceleration. Our results suggest that this interplay can be more complex than it is traditionally seen by the empirical Layzer-type buoyancy-drag models [25,39,41,47]. A direct link of the scale-dependent RT dynamics and to the self-similar RT is thus the important task. Group theory approach can be applied to solve this task, to be done in the future [11,14,22].

Our theory focuses on RT dynamics in ideal incompressible fluids. In realistic circumstances, other factors should be accounted for, including, e.g., viscosity, compressibility, surface tension, magnetic field. Qualitative and quantitative influences of these factors should be systematically studied.

For instance, viscosity may influence the instability growth-rate in the linear regime as well as the realization of asymptotic solutions in the nonlinear regime. Particularly, while the Layzer-drag bubble has a smaller velocity when compared to the Atwood bubble, it also has a larger shear and may thus 'survive' under the viscous effects, Figure 5. Likewise, viscosity may 'stabilize' for a finite time and realize the nonlinear spike with asymptotic velocity $|v| \sim \sqrt{G\lambda t^{a}}$, before the spike transits to the self-similar mixing, Figure 7.

**Sub-section 3.5.9 – Theory benchmarks**

Our analysis elaborates qualitative and quantitative benchmarks of RT dynamics not discussed before. These are the fields of velocity and pressure, the interface morphology and the curvatures of the bubble and spike, the interfacial shear and its link to the velocity and the curvature of the bubble and the spike, the spectral properties of the velocity and pressure, as well as the interface growth and growth-rate. By diagnosing the dependence of these quantities on the density ratio, the flow symmetry, and the acceleration's exponent and strength, by identifying their universal properties, and by accurately measuring departures of data in real fluids from theoretical solutions in ideal fluids, one can further advance knowledge of RT dynamics in realistic environments, achieve better understanding of RT relevant processes, from supernovae to fusion, and improve methods of numerical modeling and experimental diagnostics of interfacial dynamics in fluids, plasmas and materials under conditions of low and in high energy densities.

**Section 4 – Discussion and conclusion**

In this work, we focused on the scale-dependent dynamics of Rayleigh-Taylor instability driven by variable acceleration with power-law time-dependence for the acceleration exponents greater than (-2), and studied the spatially periodic three-dimensional flow with hexagonal symmetry in the plane normal to the acceleration direction. We applied group theory approach to solve the boundary value problem, by expanding the flow fields with the use of irreducible representations of the group, by deriving the dynamical system from the governing equations, and by finding solutions for the system to describe dynamics of RT coherent structure of bubbles and spikes in the linear and nonlinear regimes, Eqs.(1-29), Table 1 and Table 2, Figure 1 – Figure 8.

For the early-time linear dynamics, we provide the dependence of RTI growth-rate on the acceleration parameters, and show that the positions of bubbles and spikes in RT flow are defined by the initial morphology of the interface, Eqs.(10,11), Table 1, Figure 2. For the late-time nonlinear dynamics, we find the continuous family of asymptotic solutions, directly link the interface dynamics to the interfacial shear, and scrupulously study the properties of these solutions, Eqs.(15-29), Table 2, Figure 3 – Figure 8. The essentially multi-scale and interfacial character of the coherent dynamics is demonstrated. The former can be understood by viewing RT coherent structure as a standing wave with the growing amplitude. The latter implies that RT flow has effectively no motion of the fluids away from the interface and intense motion of the fluids near the interface, with shear-driven vortical structures appearing at the interface, Figure 8. Our theory provides the unified framework for the scale-dependent RT dynamics

Eqs.(1-29). Our solutions describe in the linear and nonlinear regime the RT bubbles, which move up and concave down, and the RT spikes, which move down and concave up. We find: In the linear regime RT bubbles and RT spikes grow nearly symmetrically, Figure 2. In the nonlinear regime, the dynamics of RT bubbles is regular and the dynamics of RT spikes is singular, Figure 3. We also reveal that RT bubbles and RT spikes may further accelerate due to the dominance of the vertical length scale and thus transit from the scale-dependent to the self-similar regime. Our results are congruent with previous theoretical studies [11,12,22-48] and with available observations [13-21], and elaborate extensive theory benchmarks for future experiments and simulations.

The group theory approach finds the multiplicity of nonlinear solutions for three-dimensional coherent dynamics of RT bubbles and RT spikes with variable acceleration, and associates this multiplicity with the presence of the interfacial shear. Our approach can be further applied to find solutions for three-dimensional (3D) flows with other symmetries and for two-dimensional (2D) flows [12,37,49,56]. We address the study of these problems to the future. As a brief summary - 3D flows tend to conserve isotropy in the plane, 3D highly symmetric dynamics is universal, and the dimensional 3D-2D crossover is discontinuous [12,37,49,56]. Within the frame of group theory approach, we can also directly link the scale-dependent dynamics and the self-similar mixing, to be done in the future [11, 22].

Our work studies the dynamics of ideal incompressible fluids. Our approach can be further applied to systematically analyze the influence of surface tension, viscosity, compressibility and other factors on RT dynamics with variable acceleration. We address these studies to the future. Our work focuses on the acceleration-driven RT dynamics with the acceleration exponents $a > -2$. For $a < -2$ the dynamics is driven by the initial growth-rate and is Richtmyer-Meshkov (RM) type, whereas at $a = -2$, the transition occurs from RT type to RM type dynamics with varying acceleration strength [22,23]. We provide in the future the detailed analysis of 3D and 2D flows with various symmetries for acceleration exponents $a \le -2$ [22].

Possible applications of our theory in high energy density plasmas include supernovae remnants, inertial fusion and nano-fabrication, among others [5-9,57-60]. For instance, the explosion of core-collapse supernova is driven by a blast wave, which is a variable strong shock [5,22]. The blast-wave-induced acceleration is a power-law function of time, with the acceleration exponents $-2 < a < -1$ [50-543]. According to our results, the scale-dependent interfacial dynamics driven by such accelerations is of RT type, and it is sub-exponential in the linear regime, and is decelerating in the nonlinear regime, Figure 2, Figure 3. In the self-similar mixing, the dynamics can be Richtmyer-Meshkov type and be sub-diffusive [22]. The latter implies that the energy transport at small scales may occur via localization and trapping thus influencing the synthesis of heavy mass elements in the supernova blast [22]. Our results can thus be applied to explain the abundance of chemical elements and the richness of structures observed in supernovae remnants [5].

Our group theory approach finds that Rayleigh-Taylor dynamics can in principle be controlled, in excellent agreement with experiments in fluids and plasmas at Reynolds number up to $\sim 3.2\times 10^{6}$ [14-18,58] For inertial fusion, this suggests new opportunities for control of plasma flows by means of variable acceleration and deterministic conditions [6,14,57-60]. Particularly, in the inertial confinement fusion, it may be worth to scratch the target in order to pre-impose proper deterministic conditions and to gain better control of fluid instabilities and the interfacial mixing [6,57-60]. This may help the existing methods, that are focused on a fine polishing of ICF targets in order to fully eliminate RTI [6,57].

The results of group theory approach can also be applied in nano-fabrication, by describing the dynamics and morphology of RT unstable interface in the scale-dependent regime, and the scaling and properties as well as the sensitivity to deterministic conditions in the self-similar regime [9,57-60].

To conclude, we studied the problem of RTI with variable acceleration by applying group theory approach. We linked the interface dynamics to the interfacial shear, revealed the interfacial and multi-scale character of RT dynamics, and elaborated new theory benchmarks for future research.

**Section 5 - Acknowledgements**

SIA thanks for support the University of Western Australia, AUS (project grant 10101047); and the National Science Foundation, USA (award 1404449).

**Section 6 – Data availability**

The methods, the results and the data presented in this work are freely available to the readers in the paper and on the request from the authors.

**Section 7 – Author's contributions**

The authors contributed to this work as follows: SIA - conceptualization, formal analysis, methodology, investigation, project administration, resources, supervision, writing; KCW - investigation, preparation of figures.

## Section 8 - References

**Section 9 – Tables**

Table 1: Solutions for the early-time linear dynamics of Rayleigh-Taylor bubbles and spikes

| Bubbles | Spikes |
|---|---|
| $v\,k\,\tau > 0, \quad \frac{\zeta}{k} < 0$ | $v\,k\,\tau < 0, \quad \frac{\zeta}{k} > 0$ |
| $\frac{t-t_0}{\tau} << 1, \quad v = -\frac{4}{k}\frac{d}{dt}\frac{\zeta}{k}, \quad \frac{\zeta}{k} = C_1\sqrt{\frac{t}{\tau}}\, I_{\frac{1}{2s}}\left(\frac{\sqrt{A}}{s}\left(\frac{t}{\tau}\right)^{s}\right) + C_2\sqrt{\frac{t}{\tau}}\, I_{-\frac{1}{2s}}\left(\frac{\sqrt{A}}{s}\left(\frac{t}{\tau}\right)^{s}\right), \quad s = \frac{a+2}{2}$ | |
| $t \approx t_0, \quad v - v(t_0) \approx -4A\,\frac{(\zeta_0/k)^{-1}}{(\tau k)}\left(\frac{t_0}{\tau}\right)^{a}\left(\frac{t-t_0}{\tau}\right), \quad \zeta - \zeta(t_0) \approx -\frac{k}{4}\left(\frac{t-t_0}{\tau}\right) sgn\left(\frac{v(t_0)}{v_0}\right)$ | |

Table 2: Solutions for the late-time nonlinear dynamics of Rayleigh-Taylor bubbles and spikes

| Bubbles | Spikes |
|---|---|
| $v\,k\,\tau > 0, \quad \frac{\zeta}{k} < 0$ | $v\,k\,\tau < 0, \quad \frac{\zeta}{k} > 0$ |
| $\frac{t}{\tau} >> 1, \quad v = \frac{1}{k\tau}\left(\frac{t}{\tau}\right)^{a/2} S, \quad \zeta = \xi\, k, \quad \Gamma = \frac{1}{\tau}\left(\frac{t}{\tau}\right)^{a/2} Q, \quad \Phi_m\left(\tilde{\Phi}_m\right) = \frac{1}{k\tau}\left(\frac{t}{\tau}\right)^{a/2} \phi_m\left(\tilde{\phi}_m\right)$ | |
| $S = \pm\left(9 - 64\,\xi^2\right)\sqrt{\frac{2A\xi}{48\xi - A\left(9 + 64\xi^2\right)}}$ | |
| $Q = \frac{12\,S}{9 - 64\,\xi^2}, \quad \phi_1 = -4S\frac{1+4\,\xi}{3+8\,\xi}, \quad \phi_2 = S\frac{1+8\,\xi}{3+8\,\xi}, \quad \tilde{\phi}_1 = 4S\frac{1-4\,\xi}{3-8\,\xi}, \quad \tilde{\phi}_2 = -S\frac{1-8\,\xi}{3-8\,\xi}$ | |

**Section 10 - Figure captions and Figures**

Figure 1: Schematics of the interface perturbation in Rayleigh-Taylor instability with variable acceleration for three-dimensional flow with hexagonal symmetry in the plane normal to the direction of the acceleration. (a) The interface perturbation in three-dimensional space. (b) The spatial periods and the wave-vectors of the reciprocal lattice in case of hexagonal symmetry. (c) The periodic pattern with hexagonal symmetry in the plane normal to the acceleration direction.

Figure 2: Early-time linear dynamics of Rayleigh-Taylor instability with time-varying acceleration for some values of the exponent of the acceleration power-law. The evolution of (a) the bubble curvature, (b) the bubble velocity, (c) the spike curvature, (d) the spike velocity.

Figure 3: Late-time nonlinear dynamics of three-dimensional Rayleigh-Taylor unstable flow with variable acceleration and with hexagonal symmetry in the plane. The asymptotic nonlinear solutions in the one-parameter family represent (a) the velocity and (b) the shear of the nonlinear bubbles, and (c) the velocity and (d) the shear of the nonlinear spikes as functions on their curvatures at some Atwood numbers.

Figure 4: Properties of nonlinear solutions for Rayleigh-Taylor bubbles at some Atwood numbers. The asymptotic nonlinear solutions in the one-parameter family representing (a) the velocity versus the curvature, (b) the shear versus the curvature, (c) the velocity versus the shear, and (d) the Fourier amplitudes of the fastest solution versus the Atwood number. Rayleigh-Taylor flow is driven by the acceleration with power-law time-dependence, and has hexagonal symmetry in the plane normal to the acceleration.

Figure 5: The Atwood number dependence of the parameters of special solutions in the one-parameter family for the nonlinear Rayleigh-Taylor bubbles. The asymptotic nonlinear solutions are shown for the critical bubble, the convergence limit bubble, the Taylor bubble, the Layzer-drag bubble, and the Atwood bubble. The solutions' parameters include (a) the curvature, (b) the velocity, (c) the shear, and (d) the shear versus the velocity. Rayleigh-Taylor dynamics is driven by the acceleration with power-law time-dependence and has hexagonal symmetry in the plane normal to the acceleration

Figure 6: Properties of nonlinear solutions for Rayleigh-Taylor spikes at some Atwood numbers. The asymptotic nonlinear solutions in the one-parameter family representing (a) the velocity versus the curvature, (b) the shear versus the curvature, (c) the velocity versus the shear, and (d) the Fourier amplitudes of the fastest solution versus the Atwood number. Rayleigh-Taylor flow is driven by the acceleration with power-law time-dependence, and has hexagonal symmetry in the plane normal to the acceleration.

Figure 7: The Atwood number dependence of the parameters of special solutions in the one-parameter family for the nonlinear Rayleigh-Taylor spikes. The asymptotic nonlinear solutions are shown for the critical spike, the convergence limit spike, the Taylor spike, the Layzer-drag spike, and the Atwood spike. The solutions' parameters include (a) the curvature, (b) the velocity, (c) the shear, and (d) the shear versus the velocity. Rayleigh-Taylor dynamics is driven by the acceleration with power-law time-dependence and has hexagonal symmetry in the plane normal to the acceleration

Figure 8: Qualitative velocity fields at some Atwood number for asymptotic nonlinear solutions in Rayleigh-Taylor unstable flow driven by the acceleration with power-law time-dependence and having hexagonal symmetry in the plane normal to the acceleration. The velocity fields are shown in the laboratory frame of reference near (a) the tip of the bubble and (b) the tip of the spike. The fluids moves intensively near the interface and have effectively no motions away from the interface, with vortical structures appearing at the interface due to the interfacial shear.

Figure 1

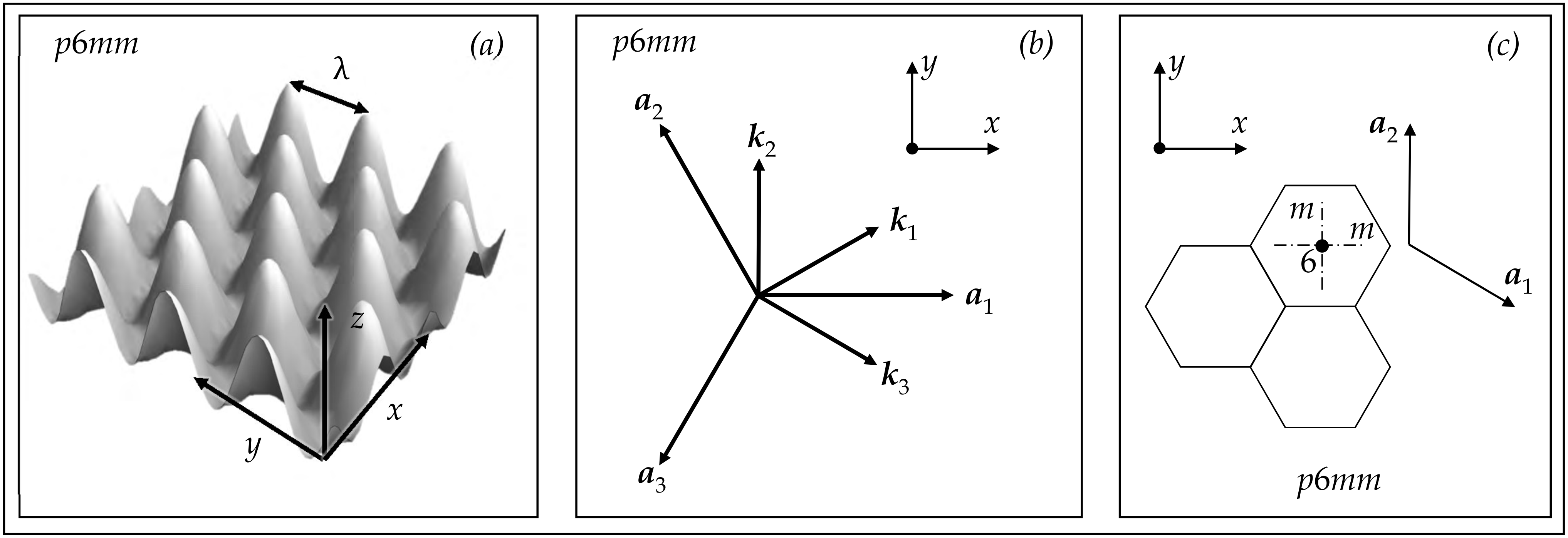

Figure 2

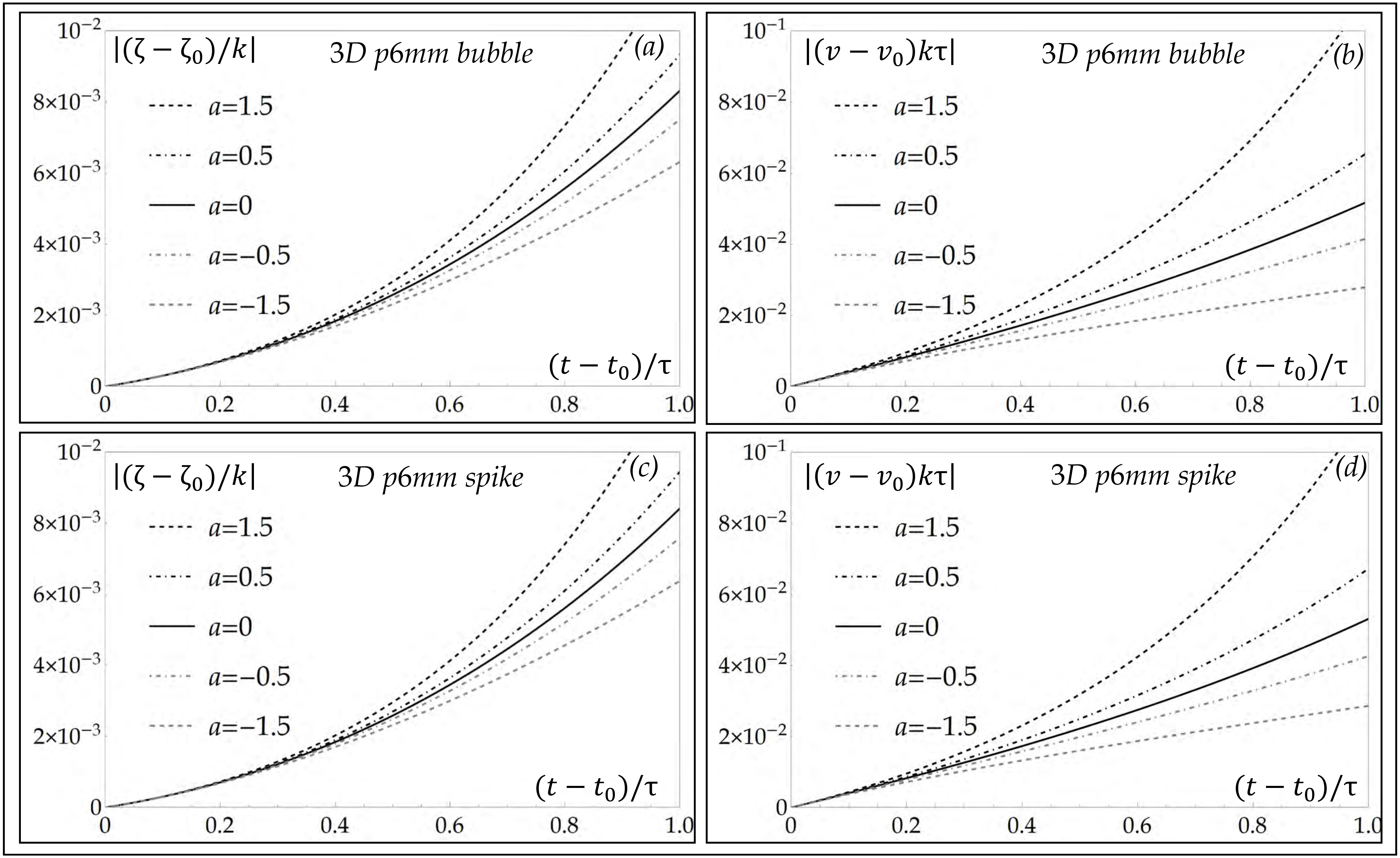

Figure 3

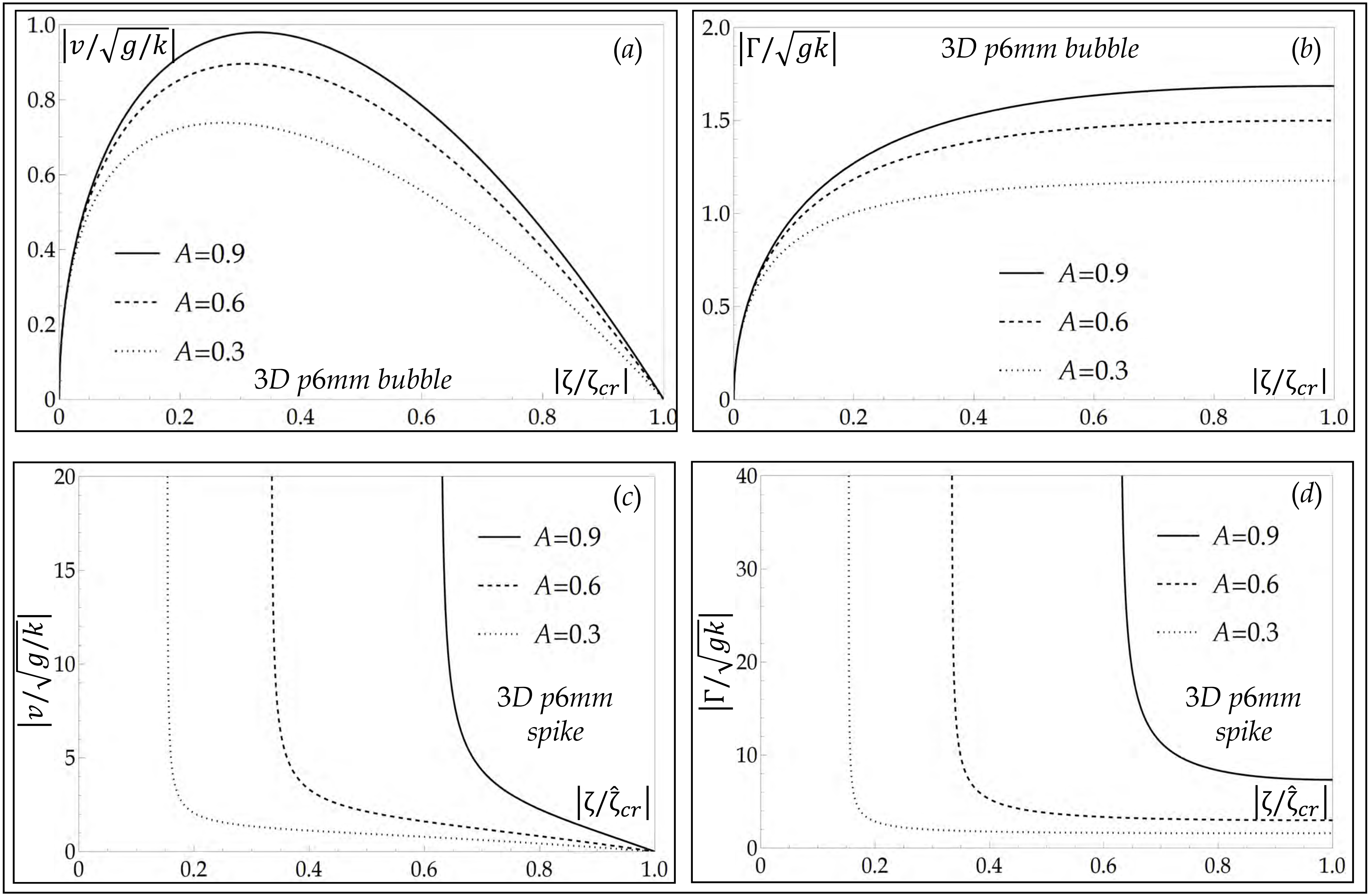

Figure 4

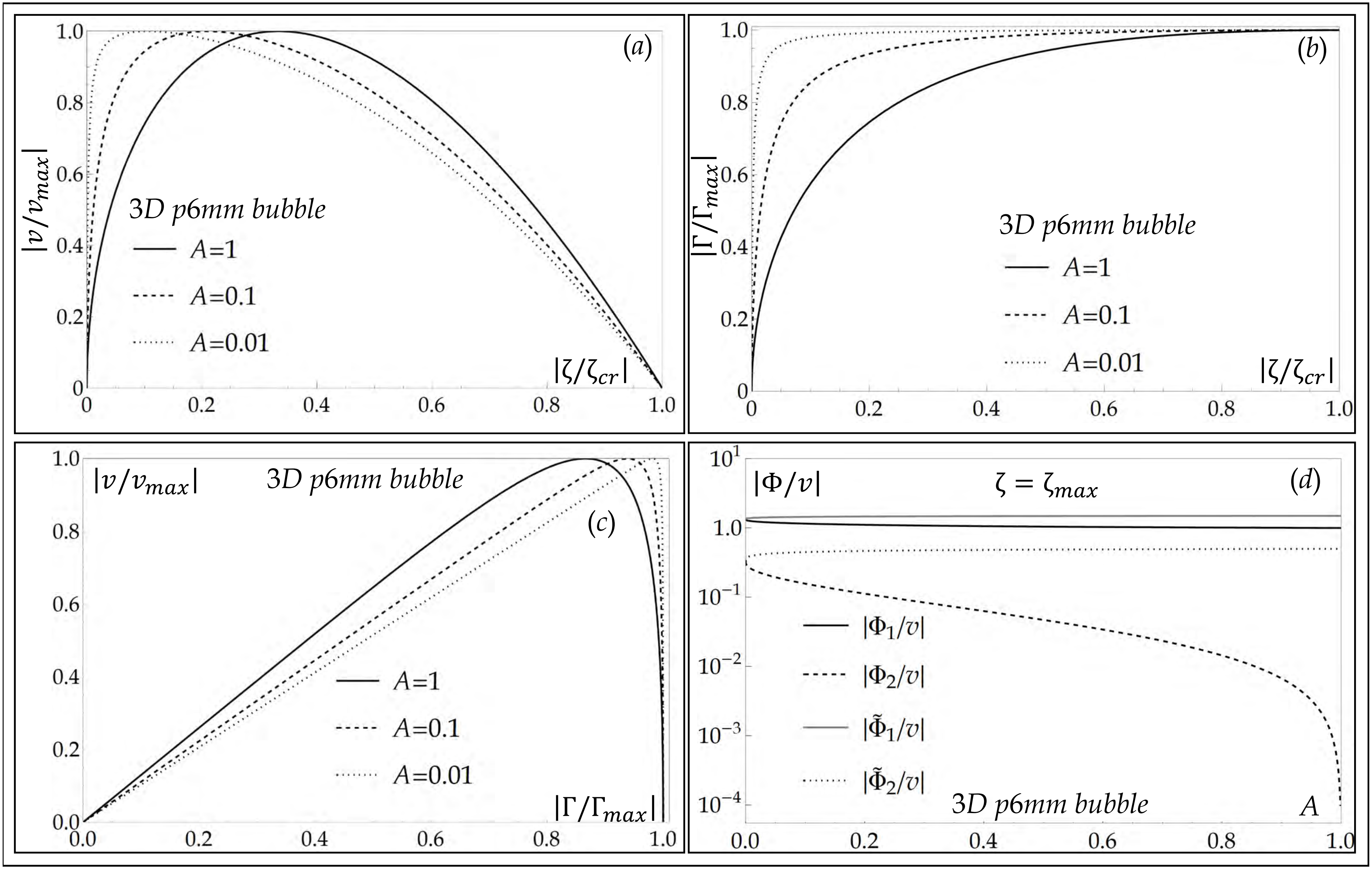

Figure 5

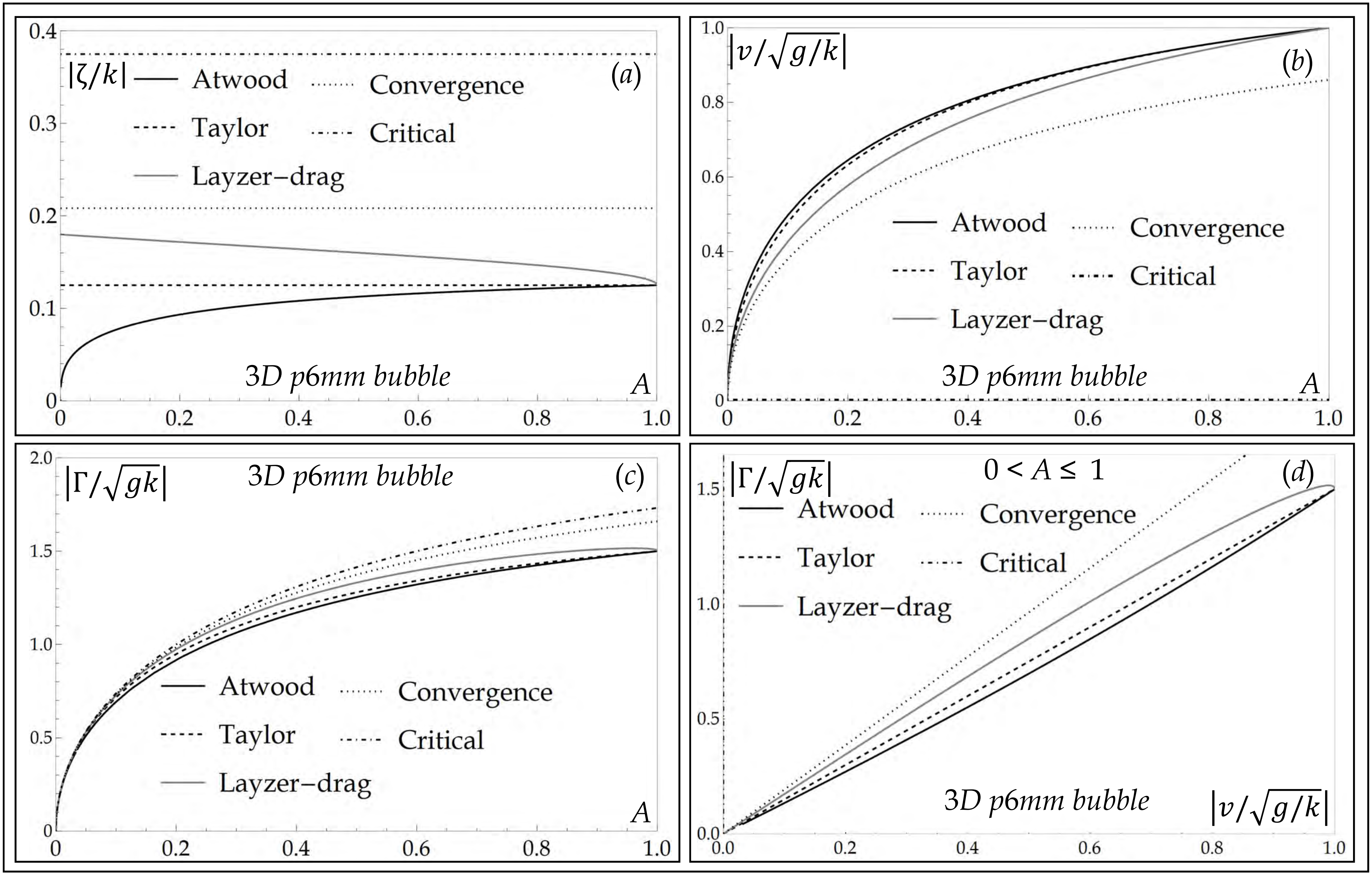

Figure 6

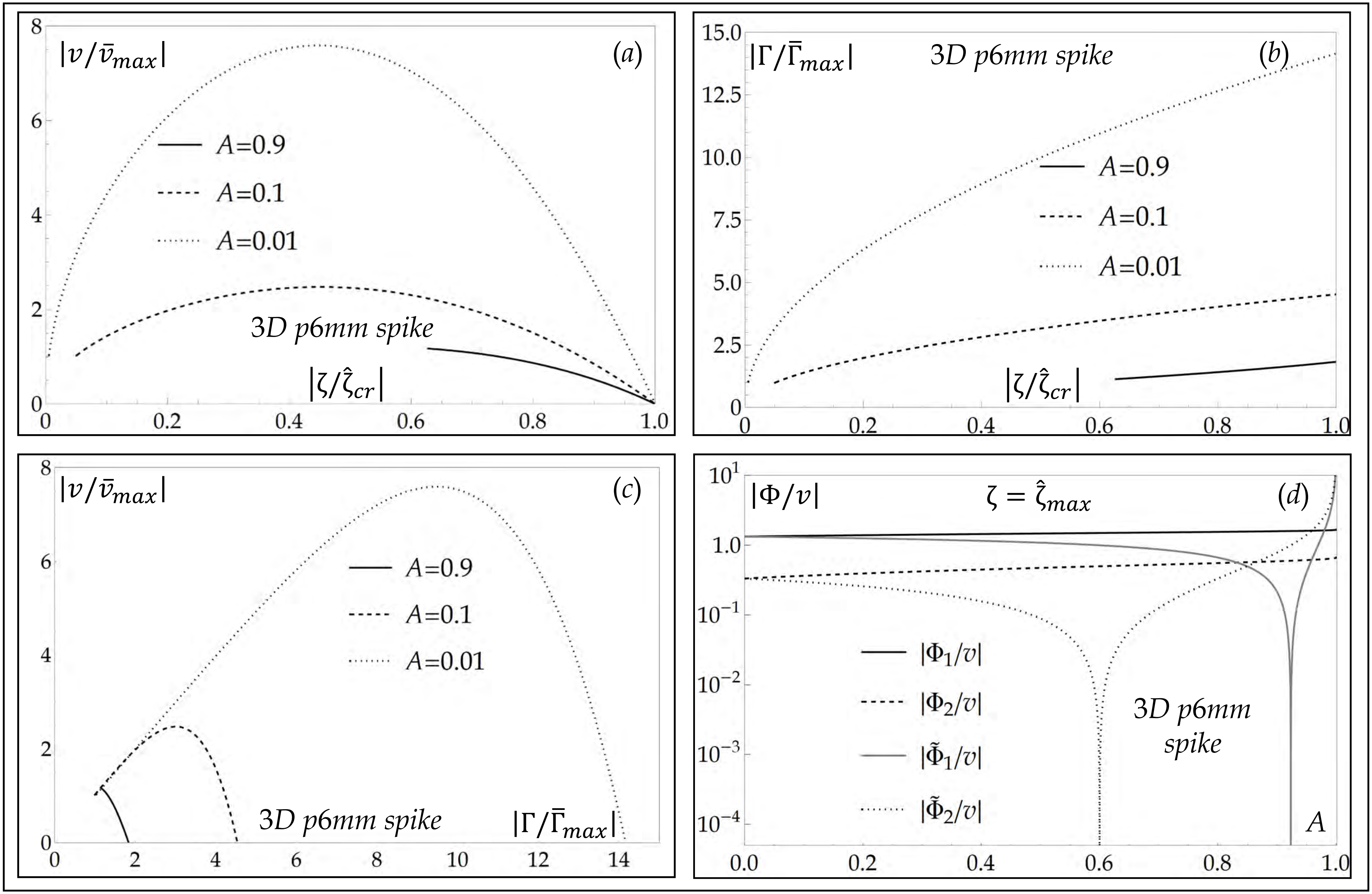

|v/v̄max|
(a)
A=0.9
A=0.1
A=0.01
3D p6mm spike
|ζ/ζ̂cr|
|Γ/Γ̄max|
3D p6mm spike
(b)
A=0.9
A=0.1
A=0.01
|ζ/ζ̂cr|
|v/v̄max|
(c)
A=0.9
A=0.1
A=0.01
3D p6mm spike
|Γ/Γ̄max|
|Φ/v|
ζ = ζ̂max
(d)
|Φ1/v|
|Φ2/v|
|Φ̃1/v|
|Φ̃2/v|
3D p6mm spike
A

Figure 7

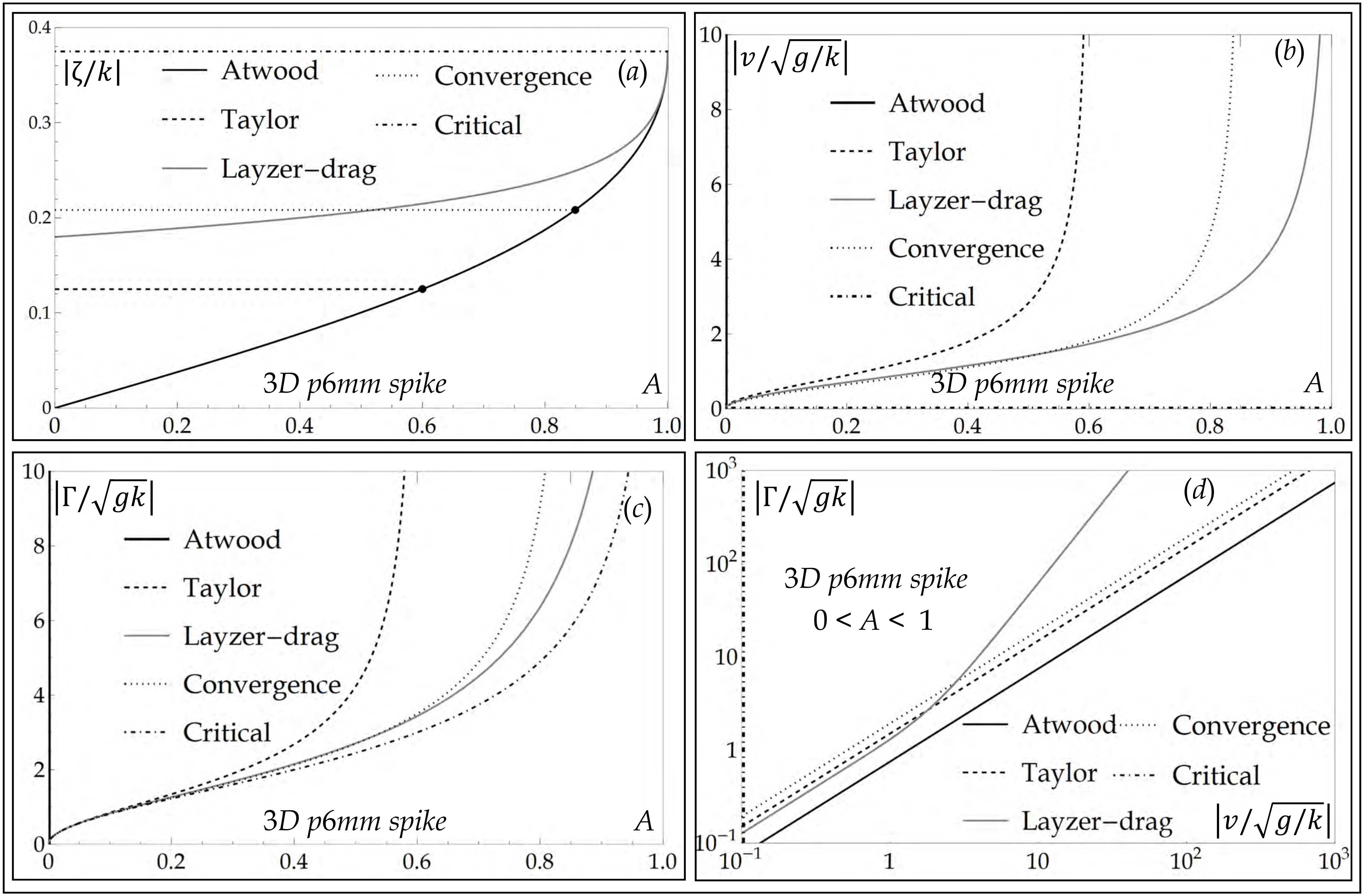

|ζ/k|
(a)
Atwood
Taylor
Layzer−drag
Convergence
Critical
3D p6mm spike
A
|v/√(g/k)|
(b)
3D p6mm spike
|Γ/√(gk)|
(c)
(d)
0 < A < 1

Figure 8

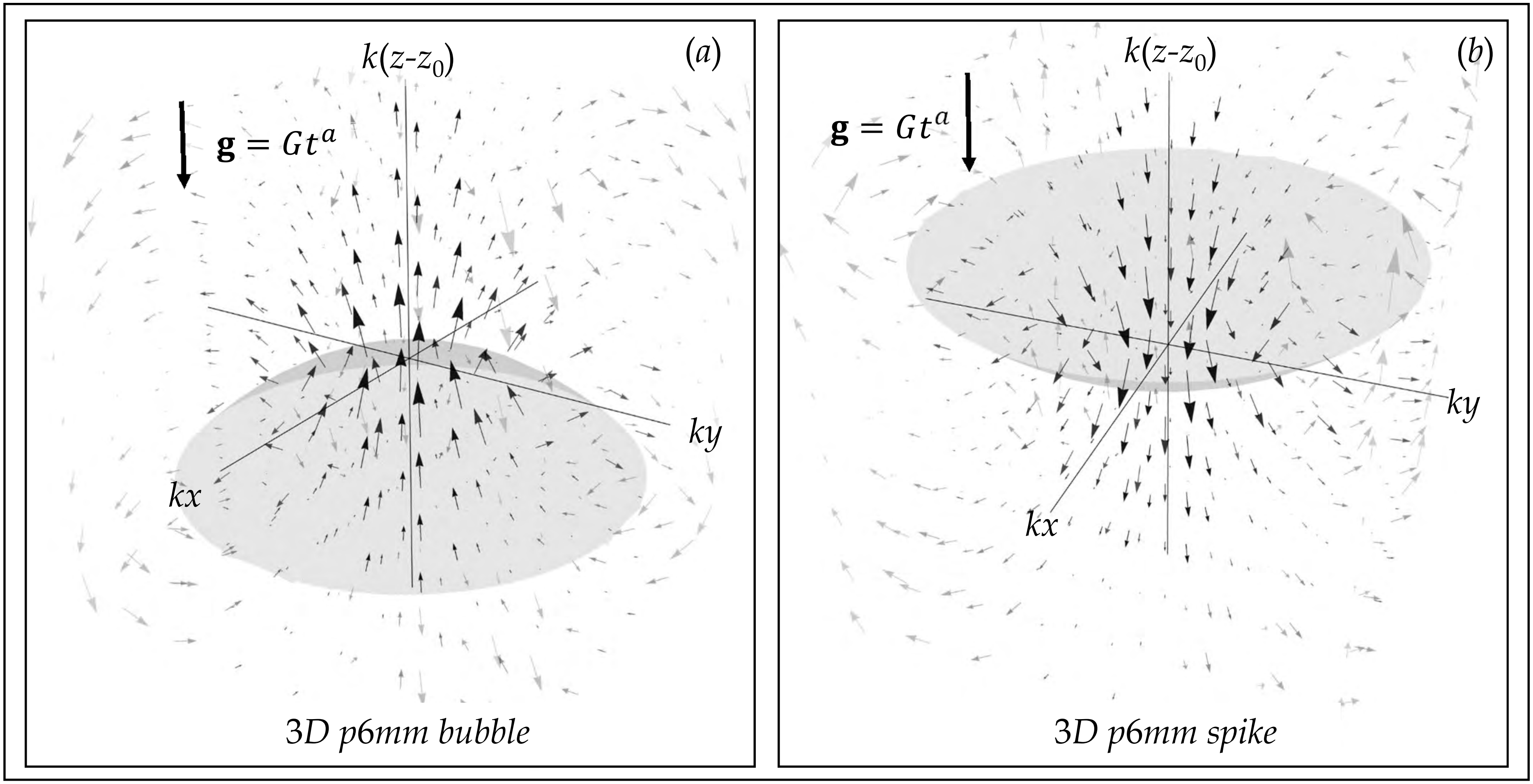